\documentclass[24pt, usenatbib]{mn2e}
\usepackage{graphicx,times}
\usepackage{bm}
\usepackage{epsf}
\usepackage{amsmath}
\usepackage{amssymb}
\usepackage{colortbl}
\usepackage{hyperref}

\def\la{\langle}
\def\ra{\rangle}

\def\be{\begin{equation}}
\def\ee{\end{equation}}
\def\ben{\begin{eqnarray}}
\def\een{\end{eqnarray}}
\def\nn{\nonumber}
\def\oh{\hat{\bf\Omega}}
\def\myC{{\cal C}}

\def\bk{{\bf k}}

\def\br{{\bf r}}

\def\bk{\bf k}
\def\bq{\bf q}

\def\myC{{\cal C}}

\def\bk{{\bf k}}

\def\2p{{(2\pi)^2}}

\def\be{\begin{equation}}
\def\ee{\end{equation}}
\def\beq{\begin{equation}}
\def\eeq{\end{equation}}
\def\ben{\begin{eqnarray}}
\def\een{\end{eqnarray}}

\def\oh{{\hat\Omega}}
\def\nn{{\nonumber}}

\def\re{{\rm e}}
\newcommand{\beqa}{\begin{eqnarray}}
\newcommand{\eeqa}{\end{eqnarray}}

\def\ex1{{\int {d^2 {\bf l} \over (2
\pi)^2}~ {\rm P}_{\pi} { \big ( {l\over d_A(r)} \big )} b_l^2(\theta_b)}}

\def\kmin1{{\int_0^{r_s}\; \omega_{\rm SZ}(r)\; dr }}

\def\ex{{\cal I}_{\theta_b}}
\def\kmin{{\Theta_m}}
\def\vp{\varphi}
\def\vt{\vartheta}
\begin{document}
\onecolumn
\title[Kinetic Sunyaev-Zel'dovich Effect]
{Extracting the late-time kinetic Sunyaev-Zel'dovich effect}
\author[Munshi et al.]
{D. Munshi, I. T. Iliev,  K. L. Dixon, P. Coles\\
Astronomy Centre, School of Mathematical and Physical Sciences, University of Sussex, Brighton BN1 9QH, U.K\\}
\maketitle
\begin{abstract}
We propose a novel technique to separate the late-time, post-reionization component of the kinetic Sunyaev-Zeldovich (kSZ) effect from the contribution to it from a (poorly understood and probably patchy) reionization history.   
The kSZ effect is one of the most promising probe of the {\em missing baryons}
in the Universe.  We study the possibility of reconstructing it in three dimensions (3D), using future spectroscopic surveys such as the Euclid survey. 
Unlike the thermal Sunyaev-Zel'dovich (tSZ) effect,
direct cross-correlation of kSZ effect with projected galaxy distribution suffers from
cancellation due to the vector (or spin-$1$) nature of the kSZ effect. By reconstructing a 3D template from
galaxy density and peculiar velocity fields from spectroscopic surveys we cross-correlate
the estimator against CMB maps. The resulting cross-correlation can help us to 
map out the kSZ contribution to CMB in 3D as a function of redshift
thereby extending previous results which use tomographic reconstruction.
This allows the separation of the late time effect from the contribution owing to reionization.
By construction, it avoids contamination from foregrounds, primary CMB, tSZ effect as well 
as from star forming galaxies. Due to a high number density of galaxies the 
signal-to-noise (S/N) for such cross-correlational studies are higher, compared to the studies involving CMB 
power spectrum analysis. Using a spherical Bessel-Fourier (sFB) transform
we introduce a pair of 3D power-spectra:  $\myC^{\parallel}_\ell(k)$  and  $\myC^{\perp}_\ell(k)$
that can be used for this purpose. We find that in a future spectroscopic survey with
near all-sky coverage and a survey depth of $z\approx 1$, reconstruction of  $\myC^{\perp}_\ell(k)$ can be achieved in a
few radial wave bands $k\approx(0.01-0.5 h^{-1}\rm Mpc)$
with a S/N of upto ${\cal O}(10)$ for angular harmonics in the range $\ell=(200-2000)$. 
\end{abstract}
\begin{keywords}: Cosmology-- kinetic Sunyaev Zel'dovich Surveys -- Methods: analytical, statistical, numerical
\end{keywords}
\section{Introduction}
Only 50\% of the baryons consistent with the cosmic microwave background radiation (CMBR) and 
big bang nucleosynthesis (BBN) observations have been detected observationally \citep{FP04,FP06}; the validation of the standard
cosmological model relies on our ability to detect the missing baryons observationally \citep{Breg07}.
Using data from the Canada France Hawaii Lensing Survey\footnote{\url http://www.cfhtlens.org/} and the Planck satellite\footnote{\url http:/www.rssd.esa.int/index.php?project=planck}, recent studies,
that use cross-correlation of the thermal Sunyaev-Zeldovich (tSZ) effect and weak gravitational lensing,
have shown that up to $50\%$ of the baryons may reside outside the virial radius of the halos \citep{MWHHS}.
The cosmological simulations too suggest that majority of the intergalactic medium (IGM) are in the form of warm-hot intergalactic medium (WHIM)
within a temperature range $10^5{\rm K}<{\rm T}<10^7{\rm K}$ \citep{CO99,Dave01,CO06}. It is also believed that WHIMs reside in moderately overdense 
structure such as filaments. However, being collisionally ionized these baryons do not leave any
footprints in Lyman-$\alpha$ absorption systems. The emission from WHIMs in either UV photons or X-rays
is too weak to be detected given the sensitivity of current instruments. Their detection is also
not feasible in X-ray given the low level of emission from WHIM. However 
the baryons in the cosmic web have sufficient velocity and column density 
to produce a detectable CMB secondary anisotropy effect known as the kinetic Sunyaev-Zel'dovich (kSZ) effect \citep{SZ80}.

CMB secondary anisotropies arise at all angular scales; the largest secondary anisotropy
at the arcminute scale is due to the thermal Sunyaev-Zel'dovich (tSZ) effect. 
The tSZ effect is caused by the thermal motion of electrons mainly
from hot ionized gas in galaxy clusters where as the {\em kinetic} Sunyaev-Zel'dovich (kSZ) effect is attributed to the
bulk motion of electrons in an ionized medium \citep{SZ72,SZ80}.
The tSZ can be separated from CMB maps using spectral information.
Along with lensing of CMB, the kSZ is the most dominant secondary
contribution at arcminute scale after the removal of tSZ, as primary CMB
is sub-dominant due to the Silk-damping \citep{OV86,Vis87,JK98,MF00,VBS01}.

The {\em thermal} and {\em kinetic} Sunyaev-Zel'dovich effects 
(tSZ and kSZ respectively) are both promising probes of the ionized fractions of the baryons.  
Majority of tSZ effect is caused by electrons in virialized collapsed objects \citep{WSP09,HH09}.
The overdensity in these regions is considerably high, $\delta >100$. The missing baryons however are more likely
to reside in filaments with moderate overdensity and lower temperature and hence unlikely to be probed
efficiently by tSZ. Interestingly, the kSZ is linked to the peculiar velocity which is sensitive 
to large scale potential and less likely to be sensitive to individual collapsed objects. 
It is also less sensitive to the thermal state and can probe contributions from individual baryonic 
components in a relatively unbiased way \citep{GAMK09}. For previous studies of kSZ effect from WHIMs see e.g. \cite{AMG08,GAMK09}.
The methods for extracting the kSZ which we present here are free from systematics that one might encounter 
when trying to detect it using the CMB two-point auto correlation function or equivalently the power-spectral analysis.

The kSZ signal is not only extremely weak; it also lacks any spectral feature \citep{MF00, MaFry02, ZPT04} and is overwhelmed 
by the primary  CMB below $\ell \ge 3000$. The kSZ effect dominate at the damping tail of the primary
beyond $\ell\ge 3000$ only at $\nu=217$~GHz where the dominant {\em secondary} contribution from the non-relativistic 
tSZ effect vanishes. Contamination from dusty star forming galaxies can also further complicate detection of
kSZ effect \citep{Hall10}. Use of non-Gaussianity has been suggested as possible method to separate kSZ from confusion due to lensing of CMB 
\citep{Castro04,RS07}. 

Being a CMB secondary anisotropy, the kSZ effect includes projected contributions from various independent components
and lacks any redshift information. It is a projection of electron momentum along the line of sight. Lack of redshift
information means it is impossible to separate individual contribution such as from a
{\em patchy} reionization history, which is expected to contribute significantly and thus can in principle
bias any estimates of missing baryons. The tight cross-correlation of tSZ and galaxy densities
has been probed in the literature in detail \citep{ZP01,Shao11btSZ}. However such a 
cross-correlation vanishes due to the vectorial (spin-1) nature of the kSZ. To circumvent this problem various options
have currently been tried, e.g. using higher order cumulant correlators \citep{DHS04,DST05} or tomographic
cross-correlation with galaxy redshift surveys \citep{Shao11a}. Cross-correlation with other tracers of low redshift large scale
structure e.g. intensity mapping from future 21cm surveys has also been considered \citep{Chang08}.

The recently proposed technique dubbed kSZ {\em tomography} \cite{HDS09, Shao11a},involves
reconstructing the galaxy peculiar velocity field from spectroscopic redshift
surveys and including galaxy number density and other redshift
dependent pre-factors. Notice that bulk flows on scales of $\ge 1{\rm Gpc}$ and the
pairwise velocity dispersion have been observed \citep{Kash08,Kash10,Kash11,Hand12,Mork12}.
The reconstructed momentum field has very similar
directional dependence as the true kSZ effect. Thus the cross-correlation of these
two can be used as an indicator for the kSZ contribution to CMB maps.
With redshift space binning it is also possible to study evolution of
kSZ contribution as a function of redshift. This method can be useful
in separating out not only the primary CMB from kSZ but also various
contribution such as the contribution from patchy reionization.
A high value for the signal to noise $({\rm S}/{\rm N})=50$ was estimated using Planck\footnote{\url http://www.cosmos.esa.int/web/planck} 
and the BigBoss\footnote{\url http://www.bigboss.lbl.org.gov/index.html} survey.
Notice that similar tomographic reconstruction was also proposed for tSZ \citep{Shao11btSZ} as was recently extended to 3D recently \citep{PM14}.

In recent years there has been significant progress in extracting survey information using three-dimensional
spherical Fourier-Bessel (sFB) decomposition of galaxy surveys which goes beyond a tomographic treatment \citep{HT95,BHT95, PM12,PM13}.
The formalism involves expanding the galaxy density field in a sFB
expansion. The formalism was found extremely useful in analysing weak lensing data in 3D \citep{H03,Castro05,MuKitch11}.
Going beyond the tomographic cross-correlation in this paper we use a 3D treatment to reconstruct the kSZ from galaxy surveys.
This is in line with the reconstruction technique developed for tSZ by \cite{PM14} using a 3D sFB decomposition.

Spectroscopic galaxy redshift surveys, such as LAMOST\footnote{\url http://www.lamost.org/website/en}, 
BOSS\footnote{\url http://www.cosmology.lbl.org.gov/BOSS}, BigBOSS, 
SKA\footnote{\url http://www.skatelescope.org}, Euclid\footnote{\url http://www.sci.esa.int/euclid} and 
JDEM/ADEPT\footnote{\url http://www.jdem.gsfc.nasa.gov/} will map
the galaxy distribution in 3D. These surveys will have overlap in terms of observed regions of the sky
with ongoing CMB surveys, e.g., ACT \footnote{\url http://www.physics.princeton.edu/act}, SPT \footnote{\url http://www.southpole.uchicago.edu} 
and Planck. It is possible to recover the velocity  fields
from such surveys and construct an estimator in 3D that can be correlated with CMB maps to reconstruct the
distribution of ionized gas whose bulk velocity is causing the kSZ. 

The primary aim of this paper is to disentangle the low-redshift (late-time) contribution to kSZ effect
and that from a rather poorly understood reionization history. This will allow us to
to use future kSZ surveys in constraining the interaction of Dark Energy (DE) and Dark Matter (DM)
using the kSZ effect as such a non-gravitational coupling in the
dark sector can significantly affect the expansion history of the Universe as well as
growth of structure \citep{Xu13}. Indeed various modified gravity theories
that affect the growth rate and power spectrum of perturbation can also
be constrained using kSZ effect.
The results presented here will be relevant to extending
studies of kSZ in constarining radial inhomegeneites in Lemaitre-Tolman Bondi cosmologies \citep{BCF11}
or dark flow \citep{Zhang10}.

The current upper limit on the KSZ power spectrum from the Atacama Cosmology Telescope (ACT) is $8.6\mu \rm K^2$
at $\ell=3000$.  The upper limits from the South Pole Telescope (SPT) is  $2.8\mu \rm K^2$ at $\ell=3000$. Most recently \citet{George15}
reported ${\cal D}^{\rm kSZ}_{\ell=3000}=(2.9\pm 1.3) \mu K^2$ (${\cal D}^{\rm kSZ}_\ell=\ell(\ell+1){\cal C}^{\rm kSZ}_{\ell}/2\pi$) .
These upper limits depend on the reionization history, modelling of Cosmic Infra-red Background, tSZ contribution
and separation of the kSZ effect into homogeneous and patchy components.
Using arcminute resolution
maps from ACT and BOSS statistical evidence was presented for motion of galaxy clusters through their kSZ signature \citep{Hand12}. 
The foreground-cleaned maps from Planck nominal mission data were used to achieve a $1.8-2.5\sigma$ detections of the kSZ signal.
This was achieved by estimating the pairwise momentum of the kSZ temperature fluctuations at the positions of the 
CGC (Central Galaxy Catalogue) samples extracted from Sloan Digital Sky Survey (DR7) data \citep{PlanckSZ15}.
In the same study, data from WMAP\footnote{\url http://map.gsfc.nasa.gov/} i.e. WMAP-9yr W band data a detection of $3.3\sigma$ was also reported.

The best constraints on reionization history are likely to result from redshifted 21-cm experiments
with low-frequency radio interferometers 
GMRT\footnote{\url http://gmrt.ncra.tifr.res.in/}\citep{Paciga11},
LOFAR\footnote{\url http://www.lofar.org/}\citep{Harker10}, 
MWA\footnote{\url http://www.mwatelescope.org/}\citep{Lonsdale09}
and PAPER\citep{Parson10}.
The near infra-red background measurements with 
CIBER\footnote{\url http://physics.ucsd.edu/ bkeating/CIBER.html} 
and AKARI\footnote{\url http://irsa.ipac.caltech.edu/Missions/akari.html} will also provide valuable clues to
reionization history. In addition, ground-based observations of high-redshift QSO, galaxies and GRBs
too provide important clues to reionization history \citep{Ouchi10,Kashikawa11,Mortlock11,Cucchiarra11}.

The plan of this paper is as follows. The \textsection\ref{sec:Not} is devoted to introducing our notation.
In \textsection\ref{sec:3Dest} we motivate a 3D analysis for kSZ reconstruction. 
In \textsection\ref{sec:3Dps} derive the expressions for 3D power spectrum. 
In \textsection\ref{sec:tomo} we present how to construct the tomographic and 2D maps and power spectrum
from 3D information. The section \textsection\ref{sec:s2n} gives analytical expression
for S/N of 3D reconstruction. The section \textsection\ref{sec:ps} is devoted to the discussion of the 
two power-spectra $P_{q\perp}$ and $P_{q\parallel}$ which are used as an input for the kSZ power spectra computation.
The \textsection\ref{sec:results} is reserved for discussion of our results. Finally, the conclusions are presented in \textsection\ref{sec:conclu}.
We have included the discussions of issues related to the partial sky coverage in Appendix-\ref{sec:mask}. 
\section{A Note on Notations} 
\label{sec:Not}
%
In this section we introduce our cosmological notations for the kSZ effect that will be used later. We will use the following line element:
\ben
&& ds^2 = -c^2 dt^2 + a^2(r)(dr^2 + d^2_{\rm A}(r)(\sin^2\vt d\vt^2 + d\vp^2) )
\een
Throughout $c$ will denote speed of light and will be set to unity.
In our notation, $d_{\rm A}(r)$ is the comoving angular diameter distance at a (comoving) radial distance $r$ and 
can be expressed in terms of the curvature density parameter $\Omega_{\rm K} = 1 - \Omega_{\rm M} - \Omega_{\Lambda}$.
Here ${\rm H}_0=100\; h \;{\rm \; km \; sec^{-1}\; Mpc^{-1}}$ is the Hubble constant; $\Omega_{\rm M}$ 
is the current non-relativistic matter density in units 
of critical density and $\Omega_{\rm \Lambda}$ is current contribution of cosmological constant to the critical density.
The angular diameter distance will be denoted as $d_A(r)$ as a function of the radial comoving distance $r(z)$ at a redshift $z$:
\ben
&& d_{\rm A} \left( r \right) =  {\lambda_{\rm H}}\frac{\sin_{\rm K} ( | \Omega_{\rm K} |^{1/2} r / \lambda_{\rm H} )}
{ | \Omega_{\rm K} |^{1/2} }; \quad  \lambda_{\rm H} = {c}{\,H^{-1}_0};
\een
where $\sin_{\rm K}$ means $\sinh$ if $ \Omega_{\rm K} > 0$ or $\sin$ if $ \Omega_{\rm K} < 0$;\;\; if $\Omega_{\rm K} = 0$,\; then $d_A \left( r \right) = r$.
The radial comoving distance $r(z)$ from a source at redshift $z$ to an observer at $z = 0$ is given by:
\ben
&& r(z) = \lambda_{\rm H}\displaystyle\int\limits^z_0 \frac{dz^{\prime}}{{E} (z')}; \quad\quad 
{E}(z) = \frac{{H}(z)}{{H}_0} = \sqrt{\Omega_{\Lambda} + \Omega_{\rm K} \left( 1 + z \right)^2 + \Omega_{M} \left( 1 + z \right)^3}. 
\label{eqn:Ez}
\een
We will denote the linear growth factor as $D_{+}(r)$ defined such that the 
Fourier transform of the over-density field grows as $\delta({\bf k},r)=D_{+}(r)\delta({\bf k},0)$ is given by:
\ben
&& D_{+}(z) = \frac{H(z)}{H_0} \int_{z}^{\infty} dz^{\prime} (1+z^{\prime}) \left [H(z') \right ]^{-3}\Big /
 \int_{0}^{\infty} dz^{\prime\prime} (1+z^{\prime\prime}) \left [H(z^{\prime\prime}) \right ]^{-3}.
\een

The particular cosmology that we will adopt for numerical 
study is specified by the following parameter values (to be introduced later):
$\Omega_\Lambda = 0.741,\; h=0.72,\; \Omega_b = 0.044,\; \Omega_{\rm CDM} = 0.215,\;
\Omega_{\rm M} = \Omega_b+\Omega_{\rm CDM},\; n_s = 0.964,\; w_0 = -1,\; w_a = 0,\;
\sigma_8 = 0.803,\; \Omega_\nu = 0$.   
%
\section{Late-time kSZ}
\label{sec:3Dest}
The cosmic ionized transverse momentum can be studied as a temperature fluctuation in the CMB maps.
The line-of-sight (los) component of the momentum of electrons in the ionized
IGM introduces temperature fluctuation via the Doppler effect also known as kSZ.
The contribution from the longitudinal component suffers cancellation due to 
contribution from successive troughs and crests \citep{Vis87}.
The measured kSZ power spectrum is the sum over contribution from the epoch of reionization (EoR)
and the post-reionization or late-time kSZ contribution. Inhomogeneity in the ionization fraction
during EoR gives a large boost to the kSZ power spectrum. The physics that determines the 
{\em patchy} or inhomogeneous reionization is complex. 
Analytical \citep{GruHu98,Santos03}, semi-analytical \citep{Zahn11,McQuinn05, Mesinger12,Zahn11,Batta13} 
and numerical N-body simulations coupled with radiative transfer \citep{Iliev07,Park13} are often used to understand various aspects.
The contribution from {\em late-time} kSZ is relatively easier to model and will be the main focus of this paper.

Inhomogeneity in electron density, defined by the density contrast $\delta_\re({\bf r})$ and peculiar velocity ${\bf v}({\bf r})$
as well as in ionization fraction $\chi_\re$ can all lead to secondary fluctuation in the CMB temperature 
$\Theta({\bf\oh}) \equiv{\delta {T}({{\bf \oh}})/{T_0}}$.
The specific ionized momentum field ${\bf q}({\bf r})$
at ${\bf r}= (r,{\bf\oh})= (r,\vt,\vp)$ is given by \citep{GruHu98,KnoxSocDod98}:
\ben
&& {\bf q}({\bf r}) \equiv \chi_\re(r) {\bf v}({\bf r}) (1+ \delta_\re({\bf r}))
\label{eq:momentum}
\een
The ionized fraction $\chi_\re$ is defined in terms of electron density $n_\re$ and number density of Hydrogen(H)
and Helium(He) atoms denoted as $n_{\rm H}$ and $n_{\rm He}$ respectively i.e. $\chi_\re = {n_\re /(n_{\rm H} + 2n_{\rm He})}$.
For post reionization or late time kSZ computation we will set $\chi_\re=1$.
If we assume that hydrogen reionization was finished at $z=6$ and HeII reionization occurred instantaneously at $z=3$
we will have $\chi_\re=0.93$ for $3<z<6$ and $\chi_\re=1$ for $z<3$. Our reconstruction technique is based on
local large scale tracers which typically cover the redshift range $z<3$ which justifies our choice of $\chi_\re=1$. 
Modelling of the post-reionization signal is simpler because the IGM is fully ionized 
and $\chi_\re$ does not fluctuate to a good approximation: ${\bf q}({\bf r}) \equiv {\bf v}({\bf r}) (1+ \delta_\re({\bf r}))$.
We thus need to model the electron density $\delta_\re$ and velocity ${\bf v}$ fluctuation of gas.
The velocity filed is purely longitudinal in the quasi-linear regime.
In general the baryonic density contrast $\delta_\re({\bf r})$ can be different from the dark matter density especially on scales 
smaller than the Jeans-length scale. Indeed the shock-heated gas will be less clustered than the underlying
dark-matter due to baryonic pressure. This effect can have modest impact on kSZ power spectrum at $\ell=3000$.
Star formation converts gas into stars thus reducing the kSZ effect further; gas-cooling and star formation
can reduce the kSZ effect up to $33\%$. These effects can only be dealt with using numerical simulations.
For this paper we will assume that the baryons trace the dark matter
on the scales of interest which are much larger than the Jeans length scale.
   
The kSZ is given by the transverse momentum field, hence is a vector mode which can be described also also as a spin-1 field. 
\ben
&& \Theta({\bf\oh})\equiv{\Delta {\rm T}({\bf \oh}) \over {\rm T_0}} =  \int_0^{z_{\rm LSS}} \; dr \, \phi(r) \Psi({\bf r}); \quad \Psi({\bf r})={{\bf\oh} \cdot {{\bf q}{({\bf r})}} \over c};\\
&& \phi(r) = -{\sigma_{\rm T} \chi_\re {\bar n}_{\re,0} \over c} {e^{-\tau(r)} \over a^2}.
\een
The projected momentum $\Psi({\bf r})$ is the 3D momentum projected along the line of sight direction at a distance ${\bf r}$. 
Here ${\bf \oh}$ is the unit vector along the line of sight direction, $r$ is the comoving distance 
along the line of sight direction.
We also note that the optical depth $\tau$ due to Thompson scattering can be expressed as:
\ben
&& d\tau= c\, n_\re(z)\, \sigma_{\rm T}\, {dt \over dz}\,dz.
\een
 where $\tau$ is the optical depth to Thompson scattering
integrated from $z=0$ to surface of last scattering at $z\sim10^3$;  $\sigma_{\rm T}= 0.66524574\times 10^{-24}\rm cm^2$ is the Thompson scattering cross-section
for electrons and ${\bar n}_{\re,0}= {\bar n}_{\rm H} + 2{\bar n}_{\rm He}$.
The visibility function $g(t)$ is defined as: 
\ben
&& g(t) \equiv -\dot\tau \exp[-\tau(t)] = c\,\sigma_{\rm T}\, n_\re(t)\,\exp[-\tau(t)].
\label{eq:g}
\een 

In this paper, we propose to construct a 3D estimator $\hat\Theta({\bf r})$ for kSZ induced temperature fluctuation
in CMB maps. The corresponding projected $\hat \Theta({\bf \oh})$ and tomographic estimators $\hat \Theta_i({\bf \oh})$
can be computed using suitable line of sight integration. We construct ${\bf q}(\br)$ from 3D spectroscopic surveys:
\ben
 && \hat\Theta({\bf r}) =  \phi(r) \Psi({\bf r}) \quad\quad 
\hat \Theta({\bf \oh}) = \int_0^{z_{\rm LSS}} dr \; \hat \Theta({\bf r}); \quad \hat \Theta_i({\bf \oh}) = 
\int_{z_i}^{z_{i+1}} dr \; \hat{\Theta}({\bf r}).
\een
The weighted momentum field ${\bf q}({\bf r})$ is constructed to have the same directional dependence as the kSZ effect and thus should cross-correlate
with the CMB maps and help us to extract the kSZ component of CMB. Similar correlations have been studied before in tomographic
bins \citep{HDS09, Shao11a} where the radial integral was restricted to a particular bin. Our approach is to go beyond the tomographic
description and constructs the kSZ effect in 3D. The cross-correlation technique has several advantages over the auto-correlation 
technique. Due to the use of 3D template that involves matching characteristic directional dependence, it is possible to
avoid contamination from tSZ, dusty star forming galaxies, as well as from the Primary CMB.
This is rather important given the weak nature of the kSZ signal and overwhelming systematic error. 
The redshift information in our estimator also allows separating out the kSZ contributions from patchy reionization and late time kSZ effect. Typically 
the high number density of galaxy ensures a higher signal to noise ratio
for such an estimator. Similar 3D techniques was be used to reconstruct the thermal SZ effect \citep{PM14}.
\begin{figure}
\centering
\vspace{0.95cm}
{\epsfxsize=14 cm \epsfysize=5 cm {\epsfbox[37 374 573 591]{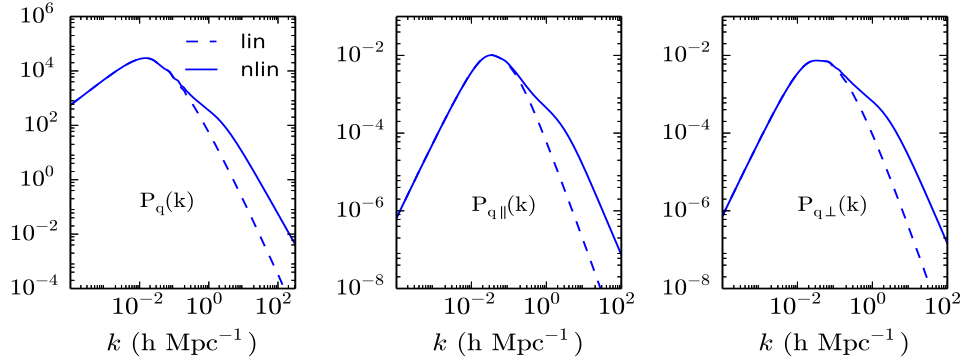}}}
\caption{The power spectrum $P_{q\parallel}(k)$ for ${\bf q}_{\parallel}$ (middle panel)  and $P_{q\perp}(k)$ (right panel)
for  ${\bf q}_{\perp}$ defined in Eq.(\ref{eq:qdef})
are plotted along with the total power spectrum  $P_q(k)=P_{q\parallel}(k)+P_{q\perp}(k)$ in the left panel as a function of $k \,(\rm h Mpc^{-1})$.
The dashed (solid) lines represent results obtained using linear (non-linear) results. The results are shown for $z=0$.}
\label{fig:pofk}
\end{figure} 
%
%
\section{3D Estimators for Late-Time kSZ}
\label{sec:3Dps}
Spherical coordinates are a natural choice for the analysis of cosmological data sets as, 
by an appropriate choice of coordinates, we can place an observer at the origin of the analysis. 
Future large scale surveys will provide an unprecedented level of detail by generating extended 
sky maps with large radial coverage. We therefore require a simultaneous treatment of the extended 
radial coverage and the spherical sky geometry. For this problem the spherical Fourier-Bessel (sFB) 
expansion is a natural basis for the analysis of cosmological random fields. The sFB formalism has 
seen an increased use in the literature over the past few years, for example it has been used in: 
weak lensing studies \citep{H03,Castro05,HKT06,MuHeCo11,Kitching11,ASW12}, redshift space distortions 
\citep{HT95,PM13,red15}, relativistic effects \citep{February13,Yoo13} and studies of baryon acoustic oscillations 
\citep{Rassat12, PM13, Grassi13}. 
 
The forward and inverse sFB decomposition of the 3D field  $\alpha(\br)$ is defined as:
\ben
&& \alpha_{\ell m}(k) = \sqrt{2 \over \pi} \int d^3{\bf r} \; k \; j_{\ell}(kr)\; Y^*_{\ell m}({\bf \oh})\; 
\alpha({\bf r}); \label{eq:forward} \\
&& \alpha({\bf r}) = \sqrt{2 \over \pi} \int kdk \; \sum_{\ell m} \; \alpha_{\ell m}(k) \;j_{\ell}(kr) Y^*_{\ell m}({\bf \oh}); 
\quad {\bf r} \equiv (r,{\bf \oh}); \label{eq:backward} \\
&& {\bf \oh}=(\cos\vartheta \sin\varphi, \sin\vartheta \sin\varphi, \cos\varphi).
\een
The basis function for the sFB expansion is an eigen function of the Laplacian operator $\sqrt{2 \over \pi} k j_{\ell}(kr)Y_{\ell m}({\bf\oh})$
in spherical co-ordinates. Here, $k$ denotes the radial wave number and $\{ \ell m \}$ denotes the azimuthal wave numbers; $Y_{\ell m}$ are the spherical
harmonics and $j_{\ell}$ are the spherical Bessel functions.
In general the continuous integral over $r$ can be a replaced with a discrete sum over galaxy position i.e.
effectively replacing the galaxy density with a sum over Dirac's delta functions $\alpha({\bf r})=\delta_{\rm D}({{\bf r-r}_i})$ that
are non-zero at galaxy position ${\bf r}_i$. Throughout results from an all-sky experiment will be presented.
For flat-sky results relevant for small patch of the sky the spherical harmonics are replaced by exponential function.

To relate the radial distance $r$ with the observed redshift of galaxies $z$ a background cosmology needs to be assumed.
The expressions derived above is technically valid for a flat Universe. To relax this assumption we have to replace the 
spherical Bessel functions with hyperspherical Bessel functions. 
To analyse realistic data weights are introduced. Such arbitrary masks will be incorporated using a 3D pseudo-${\cal C}_{\ell}$ approach.
 
Our aim is to relate the 3D harmonics of $\Theta_{\ell m}(k)$ with the underlying perturbations in specific ionized momentum field ${\bf q(r)}$ defined 
in Eq.(\ref{eq:momentum}). To this end,
we will Fourier decompose the momentum vector ${\bf q}({\bf r})$ whose Fourier components are ${\tilde\bq}(\bk)$:
\ben
&& {\tilde\bq}(\bk) = \int d^3{\bf r}\; {\bq}({\bf r})\;  e^{i{\bf k}\cdot {\bf r}}.
\label{eq:Four_q}
\een
We will separately deal with the contribution to kSZ from longitudinal ${\tilde\bq}_{\parallel}({\bf k})$ and transverse ${\tilde\bq}_{\perp}({\bf k})$ 
components of $\tilde\bq({\bf k})$ also known as the divergence free (curl) component of  ${\bq}$ i.e. ${\bq}_{\perp}$ and curl free (divergence) components.
They are defined using the following expressions:
\ben
&& {\tilde\bq}_{\parallel}(\bk) = \tilde\bq(\bk) - \tilde\bq_{\perp}(\bk); \\
&& {\tilde\bq}_{\perp}(\bk)={\bf \oh}_k[\tilde\bq(\bk)\cdot {\bf \oh}_k]; \quad\quad {\bf k} = (k, {\bf \oh}_{\rm k}).
\een
The two power spectra $P_{q_\perp}(k; r,r^{\prime})$ and $P_{q_\parallel}(k; r,r^{\prime})$ 
for ${\tilde\bq}_{\parallel}({\bf k};r)$ and ${\tilde\bq}_{\perp}({\bf k};r)$ are defined as follows: 
\ben
&& \la \tilde{\bq}_{\perp}({\bf k}; r)\tilde{\bq}_{\perp}({\bf k}'; r')\ra =
(2\pi)^3 P_{q_\perp}(k;\,r,r')\delta_{3\rm D}({\bf k}-{\bf k}'); \\
&& \la {\tilde\bq}_{\parallel}({\bf k}; r) {\tilde\bq}_{\parallel}({\bf k}'; r') \ra = (2\pi)^3 P_{q_\parallel}(k;\,r,r')\delta_{3\rm D}({\bf k}-{\bf k}').
\label{eq:qdef}
\een
The expressions for $P_{q_\perp}$ and $P_{q_\parallel}$ in terms of the matter power spectrum $P_{\delta\delta}$ will be presented in \textsection\ref{sec:ps}. 

We will next deal with the contribution to kSZ from the longitudinal and transverse components independently. First we express the
$\Theta({\bf\oh})$ in terms of the Fourier components of $\Psi({\bf r})$ i.e. $\tilde \Psi({\bf k})$. Following the same convention
for Fourier transform as described in Eq.(\ref{eq:Four_q}):
\ben
&&  \Theta({\bf\oh})= \int \; dr \; \phi(r) \int {d^3k \over (2\pi)^3}\;\;\tilde\Psi({\bf k},{\bf r})\; e^{-i\;{\bf k}\cdot{\bf r}}.
\een
We will write $\tilde\Psi = \tilde\Psi^{\parallel}+\tilde\Psi^{\perp}$ where $\tilde\Psi^{\parallel}$ depends only on ${\tilde\bq}_{\parallel}$
and $\tilde\Psi^{\perp}$ on ${\tilde\bq}_{\perp}$. Contribution to the total $\Theta({\bf\oh})$ can now be broken into two parts
namely $\Theta^{\parallel}({\bf\oh})$ and $\Theta^{\perp}({\bf\oh})$ with $\Theta({\bf\oh})= \Theta^{\parallel}({\bf\oh})+\Theta^{\perp}({\bf\oh})$.
\ben
&& \Theta^{\parallel}({\bf\oh}) \equiv \int \; dr\; \Theta^{\parallel}(\br); \; \quad\quad \Theta^{\parallel}(\br)\equiv \phi(r) \int {d^3k \over (2\pi)^3} \tilde\Psi_{\parallel}({\bf k},r) e^{-i{\bf k}\cdot {\bf r}};
\quad\quad \tilde\Psi_{\parallel}({\bf k},r) = \left [ x \; {\tilde\bq}_{\parallel}({\bf k},r) \right ];\\
&& \Theta^{\perp}({\bf\oh})=\int dr\; \Theta^{\perp}({\br}); \quad\quad\Theta^{\perp}(\br)\equiv  
\phi(r) \int {d^3k \over (2\pi)^3}\tilde\Psi_{\perp}({\bf k},r)e^{-i{\bf k}\cdot {\bf r}}; \quad
\tilde\Psi_{\perp}({\bf k},r)=\left [ \cos(\vp_q-\vp)\sqrt{(1-x^2)}\;\;{\tilde\bq}_{\perp}({\bf k},r) \right ].
\label{eq:theta_divide}
\een
We have defined $x = [{\bf \oh}_{\rm k}\cdot {\bf \oh}]$. Throughout we will denote unit vector in real space as ${\bf\oh}$ and as ${\bf\oh}_{\rm k}$ in the Fourier domain. 

Using a 3D spherical harmonic decomposition of $\Theta^{\parallel}({\br})$ and $\Theta^{\perp}({\br})$, denoted as $\Theta^{\parallel}_{\ell m}(k)$
and $\Theta^{\perp}_{\ell m}(k)$ respectively, corresponding projected (2D) harmonics $\Theta^{\parallel}_{\ell m}$ and  $\Theta^{\perp}_{\ell m}$ are obtained 
by spherical harmonics decomposition of the respective 2D analogs i.e $\Theta^{\parallel}$ and $\Theta^{\perp}$. 
We will construct two power 3D spectra $\myC^{\perp}_{\ell}(k_1,k_2)$ and $\myC^{\parallel}_{\ell}(k_1,k_2)$ similarly  to $\myC^{\perp}_{\ell}$ and $\myC^{\parallel}_{\ell}$
in 2D. We will also consider cross-correlation of our 3D harmonics with the corresponding 2D harmonics to construct $\myC^{\parallel}_{\ell}(k)$ 
and  $\myC^{\perp}_{\ell}(k)$. We will express these power spectra in terms of $P_{q_\parallel}$ and $P_{q_\perp}$.
%
\subsection{Evaluating the perpendicular components of the 3D sFB covariance ${\cal C}^{\perp}_{\ell}(k,k^{\prime})$ and the projected 2D power spectrum ${\cal C}^{\perp}_{\ell}(k)$}  
\label{sec:cperp}

In order to compute the contribution from the {\em transverse} component we will use the definition of $\Theta_{\perp}({\bf r})$ from Eq.(\ref{eq:theta_divide}) to express the power spectrum $\myC^{\perp}_{\ell}$ in terms of $P_{q\perp}$: 
\ben
&& \Theta_{\perp}({\bf r}) \equiv {\phi(r)} \int{d^3 {\bf k}' \over (2\pi)^3} \tilde\Psi_{\perp}({\bf k}',{\bf r}); \quad\\
&& \tilde\Psi^{\perp}({\bf k},{\bf r})= 
 \cos({\vp}_q-\vp)\sqrt{1-x^2}\;\;\tilde{\bq}_{\perp}({\bf k}, { r})\, e^{-i{\bf k} \cdot {\bf r}}.
\een
Next, we use the following definitions for $Y_{1\pm 1}({\bf\oh})$:
\ben
&& Y_{1,-1}({\bf\oh}) = {1 \over 2}\sqrt{3 \over 2\pi} \;\sin\vt\; e^{-i\varphi}; \quad\quad 
Y_{1,1}({\bf\oh}) = -{1 \over 2}\sqrt{3 \over 2\pi} \;\sin\vt\; e^{i\varphi}; \label{eq:sph}
\een
to express the trigonometric function in terms of lower order spherical harmonics\footnote{\url http://mathworld.wolfram.com/SphericalHarmonic.html} $Y_{\ell m}(\oh)$ :
\ben
&& \cos \vp\sin \vt = \sqrt{2\pi \over 3}[Y_{1,-1}({\bf \oh})- Y_{1,1}({\bf \oh})]. \label{eq:cossin} 
\een
We will also be using the Rayleigh expansion of plane wave in terms of spherical wave:
\ben
&& e^{-i{\bf k} \cdot {\bf r}} = 4\pi \sum_{\ell m} (-i)^\ell j_{\ell}(kr)Y_{\ell m}({\bf \oh})Y_{\ell m}({\bf \oh}_{\rm k}).
\label{eq:ray}
\een
Choosing the unit vector ${\bf\oh_{\rm k}}$ along the $z$ direction,  ${\bf\oh}_{\rm k} = \hat z$ we can replace  
$Y_{\rm \ell m}({\bf \oh}_{\rm k}) = \delta_{\rm m0} \sqrt{\Sigma_{\ell}\over 4\pi}$ with $\Sigma_{\ell}=2\ell+1$.
Using expressions from Eq.(\ref{eq:cossin}) we can write:
\ben
\tilde\Psi^{\perp}({k\hat z};{\bf r}) &=& 4\pi \sqrt{2\pi \over 3} 
[Y_{1,-1}({\bf \oh})- Y_{1,1}({\bf \oh})]\;\tilde{\bq}_{\perp}({\bf k}, r)
\sum_{\ell'm'}\,(-i)^{\ell^{\prime}}\, j_{\ell'}(kr)\,Y_{\ell' m'}({\bf \oh})\, Y_{\ell' m'}({\bf \oh}_{\rm k})\nn \\
&=& \sqrt{4\pi}\;\sqrt{2\pi \over 3}\; [Y_{1,-1}({\bf \oh})- Y_{1,1}({\bf \oh})]\; \tilde{\bq}_{\perp}({\bf k}, r)
\sum_{\ell'} \,(-i)^{\ell^{\prime}}\, \sqrt{\Sigma_{\ell'}} \;j_{\ell'}(kr) \; Y_{\ell' 0}({\bf \oh}).
\een
We will express the overlap-integrals involving three spherical harmonics in terms of $3j$ symbols:  
\ben
\tilde\Psi^{\perp}_{\ell m}({k\hat z}, r) &=&  \sqrt{4\pi}\sqrt{2\pi \over 3}
\tilde{\bq}_{\perp}({k\hat z},r)
\sum_{\ell'}\;\,(-i)^{\ell^{\prime}}\, j_{\ell'}(k\,r)\; \sqrt \Sigma_{\ell'}\;
\int \; d\oh \; Y_{\ell'0}({\bf \oh})\; [Y_{1,-1}({\bf \oh})- Y_{1,1}(\oh)] \; Y^*_{\ell m}({\bf \oh}); \\
&=&\sqrt{4\pi}\sqrt{2\pi \over 3}
\tilde{\bq}_{\perp}({k\hat z},r)
\sum_{\ell'} \,(-i)^{\ell^{\prime}}\, I_{\ell,\ell',1} \; j_{\ell'}(kr)\sqrt \Sigma_{\ell}\left(  \begin{array}{ c c c }
     \ell & \ell' & +1 \\
     0 & 0 & 0
  \end{array} \right) \nn \\
&& \quad\quad \times \left [\left(  \begin{array}{ c c c }
     \ell & \ell' & +1 \\
     -m & 0 & -1
  \end{array} \right)-
 \left(  \begin{array}{ c c c }
     \ell & \ell' & +1 \\
     -m & 0 & +1
  \end{array} \right)\right ]. \label{eq:theta_perp}
\een
The $3j$ symbols are defined through the {\em Gaunt} integral: 
\ben
&& \int \; d{\bf \oh}\; Y_{\ell_1 m_1}({\bf\oh})\; Y_{\ell_2 m_2}({\bf \oh})\; Y_{\ell_3 m_3}({\bf \oh}) = I_{\ell_1,\ell_2,\ell_3} \left(  \begin{array}{ c c c }
     \ell_1 & \ell_2 & \ell_3 \\
     0 & 0 & 0
  \end{array} \right)
\left(  \begin{array}{ c c c }
     \ell_1 & \ell_2 & \ell_3 \\
     m_1 & m_2 & m_3
  \end{array} \right);\\
&& I_{\ell_1,\ell_2,\ell_3}= \sqrt{\Sigma_{\ell_1}\Sigma_{\ell_2}\Sigma_{\ell_3} \over 4\,\pi}.
\een
For various symmetry properties of $3j$ symbols see e.g. \cite{Ed96}.
Notice, only $m=\pm1$ multipoles contribute as the $3j$ symbols need to satisfy the condition $m_1+m_2+m_3=0$. The
first term in the Eq.(\ref{eq:theta_perp}) contributes for $m=-1$ and the second term contributes for $m=1$. 

We will use the following identities in our derivation to evaluate the special cases of the 3j symbols that appear:
\ben
\left(  \begin{array}{ c c c }
     \ell_1 & \ell_2 & \ell_3 \\
     0 & 0 & 0
  \end{array} \right) &=&  {(-1)}^\ell \sqrt{(2\ell-2\ell_1)!(2\ell-2\ell_2)!(2\ell-2\ell_3)! \over (2\ell+1)!} 
{ \ell! \over (\ell-\ell_1)!(\ell-\ell_2)!(\ell-\ell_3)! }; \quad\quad \ell_1+\ell_2+\ell_3 = 2\ell \nn \\
&=& 0; \quad\quad \ell_1+\ell_2+\ell_3 = 2\ell+1.
\label{eq:3j1}
\een
\ben
\left(  \begin{array}{ c c c }
     \ell_1 & \ell_2 & \ell \\
     \ell_1 & -\ell_1 & -m 
  \end{array} \right ) = (-1)^{-\ell_1+\ell_2+m} \left [ {(2\ell_1)!(-\ell_1+\ell_2+\ell)! 
\over (\ell_1+\ell_2+\ell+1) (\ell_1-\ell_2+\ell)!} {(\ell_1+\ell_2+m)!(\ell-m)! \over (\ell_1+\ell_2-\ell)!(-\ell_1+\ell_2-m)!} \right ]^{1/2}.
\label{eq:3j2}
\een

\smallskip
\noindent
{\bf(A)}\;\; Modes with \underline{$m=-1$}
\smallskip

\noindent
We will consider the $m=-1$ term first:
\ben
\tilde\Psi^{\perp}_{\ell,m=-1}({k\hat z}; r) &=&- \sqrt{4\pi}\sqrt{2\pi \over 3}
\tilde{\bq}_{\perp}({k\hat z};r)\noindent 
\sum_{\ell'}\; I_{\ell,\ell',1} \; j_{\ell'}(k\,r) \;\sqrt{\Sigma_{\ell'}}\; \left(  \begin{array}{ c c c }
     \ell' & \ell & +1 \\
     0 & 0 & 0
  \end{array} \right) 
\left(  \begin{array}{ c c c }
     \ell' & \ell & +1 \\
     0 & +1 & -1
  \end{array} \right).
\een
The 3j symbols get contribution from $\ell'=\ell+1$ and $\ell'=\ell-1$ due to the {\em triangular} inequality:
\ben
\tilde\Psi^{\perp}_{\ell,m=-1}({k\hat z};r) &=& -\sqrt{4\pi}\sqrt{2\pi \over 3}
\tilde{\bf q}_{\perp}({k\hat z};r)
\Big [ \sqrt{\Sigma_{\ell+1}} j_{\ell+1}(k r) I_{\ell,\ell+1,1} 
\left(  \begin{array}{ c c c }
     \ell+1 & \ell & +1 \\
     0 & 0 & 0
  \end{array} \right)
\left(  \begin{array}{ c c c }
     \ell+1 & \ell & 1 \\
     0 & +1 & -1
  \end{array} \right) \nn \\
&& + \sqrt{\Sigma_{\ell-1}} j_{\ell-1}(k r) I_{\ell,\ell-1,1} 
\left(  \begin{array}{ c c c }
     \ell-1 & \ell & +1 \\
     0 & 0 & 0
  \end{array} \right)
\left(  \begin{array}{ c c c }
     \ell-1 & \ell & +1 \\
     0 & +1 & -1
  \end{array} \right) \Big ].
\een
We use the identities in Eq.(\ref{eq:3j1}) and Eq.(\ref{eq:3j2}) to derive the following expressions:
\ben
&& \left(  \begin{array}{ c c c }
     \ell+1 & \ell & +1 \\
     0 & 0 & 0
  \end{array} \right) = (-1)^{\ell +1} \sqrt{ \ell+1 \over \Sigma_{\ell} \Sigma_{\ell+1} };\quad
\left(  \begin{array}{ c c c }
     \ell+1 & \ell & +1 \\
     0 & +1 & -1
  \end{array} \right) = (-1)^{\ell -1} \sqrt{\ell\over 2\Sigma_{\ell} \Sigma_{\ell+1} }.
\label{eq:new3j1}
\een
We also use $I_{\ell,\ell\pm1,1}= \sqrt{3/4\pi}\sqrt{\Sigma_{\ell}}\sqrt{\Sigma_{\ell\pm1}}$ we eventually get:
\ben
&&\tilde\Psi_{\ell m=-1}^{\perp}({k\hat z}; r) = (-i)^{\ell+1}\;\sqrt{\pi\Pi_{\ell}\Sigma_{\ell}}  {j_{\ell}(kr) \over kr}\; 
\tilde{\bq}_{\perp}({k\hat z},r); \quad\quad \Pi_{\ell} \equiv \ell(\ell+1).
\een
Where we have also used the following relation for the spherical Bessel functions:
\ben
&& j_{\ell+1}(x)+ j_{\ell-1}(x) = {\Sigma_{\ell} \over x} j_{\ell}(x).
\een

\noindent
{\bf (B)} Modes with \underline{$m=+1$} 

\noindent
This component can also be computed in an analogous manner:
\ben
&& \tilde\Psi^{\perp}_{l, m=1}({k\hat z}; r) = 
{\tilde\bq}_{\perp}({k \hat z},r)
\sum_{\ell'} I_{\ell\ell'1} \; j_{\ell'}(kr) \; \sqrt{\Sigma_{\ell'}}\; \left(  \begin{array}{ c c c }
     \ell' & \ell & +1 \\
     0 & 0 & 0
  \end{array} \right) 
\left(  \begin{array}{ c c c }
     \ell' & \ell & +1 \\
     0 & +1 & -1
  \end{array} \right).
\label{eq:Theta_perp}
\een
Following the same steps and using the following expressions for the $3j$ symbols:
\ben
&& \left(  \begin{array}{ c c c }
     \ell-1 & \ell & +1 \\
     0 & 0 & 0
  \end{array} \right) = (-1)^{\ell} \sqrt{ \ell \over \Sigma_{\ell}\Sigma_{\ell-1} };\quad
\left(  \begin{array}{ c c c }
     \ell-1 & \ell & +1 \\
     0 & +1 & -1
  \end{array} \right) = (-1)^{\ell -1} \sqrt{\ell+1 \over 2\Sigma_{\ell}\Sigma_{\ell-1} };
\label{eq:new3j2}
\een
it can be shown that $\tilde\Psi^{\perp}_{\ell, m=1}({k\hat z};r)=-\tilde\Psi^{\perp}_{\ell, m=-1}({k\hat z}; r)$.
 
\bigskip
\noindent
{\bf (C)} Combining \underline{$m=\pm 1$} modes 
\smallskip

Next, to rotate $\Theta^{\perp}_{\ell m'}(k\hat z)$ defined for $k$ along the $z$-axis to an arbitrary observer's orientation $\Theta^{\perp}_{\ell m}(k)$, 
we use the expression $\tilde\Psi^{\perp}_{\ell m}(k) = \sum_{m'=\pm 1} D^{\ell}_{mm'}(\oh_{\bk}) \tilde\Psi^{\perp}_{\ell m'}(k\hat z)$, where $\rm D$ is the 
Wigner $\rm D$-matrix, we can write Eq.(\ref{eq:Theta_perp}) in following form:
\ben
&& \Theta^{\perp}_{\ell m}(k) = (-i)^{\ell+1}\; \sqrt{\pi\Pi_{\ell}\Sigma_{\ell}}\;
\int{d^3 {\bf k}^{\prime} \over (2\pi)^3}\; \sum_m D^{\ell}_{m,\pm1}({\bf \oh}_{\rm k^{\prime}})
\sqrt{2 \over \pi}\int dr\; \phi(r)\;k\; j_{\ell}(k\,r) {j_{\ell}(k^{\prime} r) \over k^{\prime}r}\; {\tilde\bq}_{\perp}({k}^{\prime}{\hat z};r).
\label{eq:theta_elm_perp}
\een

\smallskip
\noindent 
{\bf (D)} Constructing the power spectra 
\smallskip

Finally the 3D sFB covariance $\hat{\cal C}^{\perp}_{\ell}(k_1,k_2)$ can be expressed in terms
of  $P_{q\perp}$ as:
\ben
\;\;\;\;\;\;\;\; \hat{\cal C}^{\perp}_{\ell}(k_1,k_2)&\equiv& \la\hat\Theta_{\ell m}^{\perp}(k_1)\hat\Theta_{\ell m}^{*\perp}(k_2)\ra \nn \\
&=& 2\Pi_{\ell}\;\int \; dk\; k^2 \; \int d\,r_1\; \phi(r_1)\; k_1 j_{\ell}(k_1r_1) {j_\ell(k\,r_1) \over k r_1}
\int d\,r_2\; \phi(r_2)\; k_2\, j_{\ell}(k_2r_2)\; {j_\ell(k\,r_2) \over k r_2} \; P_{q\perp}(k; r_1,r_2).
\label{eq:full_perp}
\een
To perform the $d{\bf \Omega}_{\rm k}$ integral in Eq.(\ref{eq:theta_elm_perp}) we use the following expression for the orthogonality of Wigner $\rm D$-matrices: 
\ben
&& \int d{\bf \oh}_{\rm k} \; D^{\ell}_{m_1,m_2}({\bf \oh}_{\rm k})\; D^{\ell'}_{m_1',m_2'}({\bf \oh}_{\rm k}) = 
{4\pi \over \Sigma_{\ell}}\delta^{\rm K}_{\ell\ell'}\delta^{\rm K}_{m_1m_1'}\delta^{\rm K}_{m_2m_2'}.
\label{eq:d_matrix}
\een
We note here that even for all-sky coverage the covariance matrix above can be ill-conditioned, reflecting
the very nature of the kSZ effect which inherently non-linear even at linear regime due to
coupling of modes resulting from density and velocity field. Thus a practical implementation
may require binning to improve the condition number.
The expression presented in Eq.(\ref{eq:full_perp}) can be evaluated  directly. In case of more conservative estimates one can directly evaluate using numerical simulations. Nevertheless  we also quote results that 
use the following approximation $P_{q\perp}(k; r,r^{\prime}) \approx \sqrt{P_{q\perp}(k; r)}\sqrt{P_{q\perp}(k; r^{\prime})}$ \citep{Castro05}.
to reduce the dimensionality of the integration in Eq.(\ref{eq:full_perp}).
\ben
&& {\cal C}^{\perp}_{\ell}(k_1,k_2) \approx \int \; k^2 \; dk \; {\cal I}^{2\perp}_{\ell}(k,k_1) \; {\cal I}^{2\perp}_{\ell}(k,k_2); \label{eq:perp_cl1}\\
&& {\cal I}^{2\perp}_{\ell}(k,k') \approx  \sqrt{2 \Pi_{\ell}}
\int \;dr\; \phi(r)\; k\; j_{\ell}(kr)\;{j_{\ell}(k^{\prime}r)\over k^{\prime} r}\; \sqrt{P_{q\perp}(k^{\prime}; r)}. \label{eq:perp_cl2}
\een
The kernel ${\cal I}^{\perp}_{\ell}(k,k')$ defined above encapsulates the information regarding the mode-mode coupling.
Finally we recover the projected (2D) power spectrum for the kSZ effect is given by \cite{MaFry02}:
\ben
&&\myC^{\perp}_\ell = {1 \over 2}\int {dr \over r^2} \; \phi^2(r) \;
P_{q\perp}\left ({k= {\ell \over r}}; r \right).
\label{eq:2DCls}
\een
The derivation of Eq.(\ref{eq:2DCls}) depends on the use of Limber approximation stated below for an arbitrary function $F$:
\ben
&& \int \;k^2\; dk\; F(k) j_{\ell}(kr)j_{\ell}(kr^{\prime}) \approx {\pi \over 2 r^2}F\left (k ={\ell \over r}\right )\delta_{D}(r-r^{\prime});
\een
which generalises the orthogonality property of the spherical Bessel functions:
\ben
&& \int \;k^2\; dk\; j_{\ell}(kr)j_{\ell}(kr^{\prime}) = {\pi \over 2 r^2}\delta_{D}(r-r^{\prime}).
\een
%
Cross-correlating the CMB maps $\Theta({\bf\oh})= \sum_{\ell m}\Theta_{\ell m} Y^*_{\ell m}(\oh)$ against reconstructed 3D estimators:
\ben
&& {\hat{\cal C}}^{\perp}_{\ell}(k) \equiv \la\Theta_{\ell m}\hat\Theta_{\ell m}^{*\perp}(k)\ra = 
2\Pi_{\ell}\;\int \; dk^{\prime}\; {k^{\prime}}^2 \; \int d\,r_1\;r_1^2\; \phi(r_1)\; {j_\ell(k^{\prime}\,r_1) \over k^{\prime} r_1}
\int d\,r_2\;r_2^2\; \phi(r_2)\; k\, j_{\ell}(k\,r_2)\; {j_\ell(k^{\prime}\,r_2) \over k^{\prime} r_2} \; P_{q\perp}(k^{\prime}; r_1,r_2).
\label{eq:full_recon_perp_est}
\een
Notice that the estimator $\hat\Theta({\bf r})$ is constructed using 3D distribution of galaxies and the recovered
3D velocity distribution. The weight $\phi(r)$ included in the construction depends on a prior choice of cosmology.   
The sFB expansion decomposes the reconstructed kSZ estimator to mode specified by $\{k\ell m\}$. 
The cross-correlation of this estimator with CMB maps $\Theta({\bf\oh})$ thus extracts only kSZ contribution
as a function of radial harmonics $k$ and radial harmonics $\ell$. This provides an unique method
to recover radial information about evolution of kSZ effect. The construction has other advantages.
It avoids any contamination from other CMB secondary anisotropies such as lensing or tSZ, or even from primary CMB
and thus reduces the statistical error.
Given that the kSZ signal is weak and foreground contamination (as well as confusion from dusty star forming galaxies) 
is high at angular scales where kSZ peaks the method presented here offers an unique opportunity.

The estimator in Eq.~(\ref{eq:full_recon_perp}) involves a 3D integral.
We can use the same factorization of unequal time correlator introduced above to reduce it's dimensionality:
\ben
&& {\hat{\cal C}}^{\perp}_{\ell}(k) \approx \int \; dk\; {k^{\prime}}^2 \; {\cal I}^{2\perp}_{\ell}(k,k^{\prime})\; {\cal I}^{\perp}_{\ell}(k^{\prime});
\label{eq:factor_recon_perp1} \\
&& {\cal I}^{\perp}_{\ell}(k) \approx \int\;dr r_1^2 \;\phi(r)\; {j_{\ell}(kr)\over kr} \sqrt{P_{q\perp}(k;r)}. 
\label{eq:factor_recon_perp2}
\een
Eq.~(\ref{eq:full_recon_perp_est}) and its simplified version Eq.~(\ref{eq:factor_recon_perp1})-Eq.~(\ref{eq:factor_recon_perp2}) constitute
some of the main results of this paper.

\smallskip
\noindent
{\bf (E)} Reconstructing the velocity field: 

The construction of the estimator introduced above for obtaining a 3D galaxy distribution from a 3D survey is just a first step. The galaxy density field $\delta_g$ is used as a proxy for the underlying dark matter distribution $\delta$. It is possible to incorporate
redshift- and scale-dependent bias but at large scales where we will use
our estimator, a simple linear bias $\delta_g= b\delta$ should be sufficient.  Construction
of a kSZ estimator should include appropriate weighting using the recovered velocity field ${\bf v}$.
This is required to avoid the cancellation of cross-correlation of density field and
an estimator for kSZ due to the spinorial nature of the kSZ field. 
The reconstruction of 3D velocity field uses the linearised continuity equation.
\ben
&& \dot \delta + \nabla\cdot{\bf v}=0.
\een
In Fourier space it takes the following form:
\ben
&&\hat{\bf v}({\bf k}) = -if(z) H(z) \delta({\bf k}) {{\hat {\bf k}} \over k}; \quad\quad f(z) \equiv {d\ln D_{+} \over d\ln a}.
\label{eq:lin_den_vel}
\een
The estimator thus constructed using linear theory will be insensitive to multi-streaming resulting from shell-crossing
that introduces vorticity. However, non-linearities at the perturbative level can be included in a more sophisticated
treatment which we ignore here.  The linear redshift-space distortion (the Kaiser effect) is ignored here but 
can be incorporated in our treatment \citep{red15}. A prior on background cosmology is required for specification of
$H(z)$ and $f(z)$. 
\subsection{Ealuating the parallel components of the 3D sFB covariance ${\cal C}^{\parallel}_{\ell}(k,k^{\prime})$ and the projected 2D power spectrum ${\cal C}^{\parallel}_{\ell}(k)$}
\label{sec:cpara}
The contribution from ${\cal C}^{\parallel}_{\ell}(k)$ can be used as a check on systematics.
The computation of the {\em longitudinal} contribution or $\myC^{\parallel}_{\ell}(k_1,k_2)$ in terms of $P_{q\parallel}$ follows similar steps.
\ben
&& \Theta^{\parallel}({\bf r}) = {\phi(r)} \int{d^3 {\bf k^{\prime}} \over (2\pi)^3} \tilde\Psi_{\parallel}({\bf k}',{\bf r}); \\
&& \tilde\Psi^{\parallel}({\bf k}; {\bf r}) = 
\; \cos\vt \; \tilde{\bq}_{\parallel}({\bf k}; {\bf r})e^{-i {\bf k} \cdot {\bf r}}
\een
Using the fact:
\ben 
&& \cos \vt =\sqrt{4\pi \over 3}\;\;Y_{1,0}({\bf \oh}); \label{eq:rj}
\een
and taking advantage of Rayleigh expansion  Eq.~(\ref{eq:ray}) we can write:
\ben
&& \tilde\Psi^{\parallel}({\bf k}; {\bf r}) =  4\pi \; \sqrt {4\pi \over 3}\; \tilde{\bq}_{\parallel}({\bf k}, r) \sum_{\ell m} (-1)^{\ell} j_{\ell}(k r) 
Y_{\ell m}({\bf \oh})Y_{\ell m}({\bf \oh}_{\rm k}).
\een
Next, aligning the ${\bf k}$ vector along the $z$-axis (${\bf k}=k\,\hat z$) we use the identity $Y_{\ell m}({\bf \oh}_{\rm k})= \delta_{m0}\sqrt{\Sigma_{\ell}\over 4\pi}$:
\ben
&& \tilde\Psi^{\parallel}({k\hat z}; {\bf r}) =  4\pi \sqrt {1 \over 3}\; \tilde{\bq}_{\parallel}({k\hat z}, r)Y_{10}(\oh) \sum_{\ell'}
\sqrt{\Sigma_{\ell'}}j_{\ell'}(kr)Y_{\ell'0}(\oh).
\een
We perform a spherical harmonic transform of $\tilde\Psi^{\parallel}$.  We note that the only non-zero harmonic coefficients come from \underline{$m=0$}. Using the expression $I_{\ell,\ell\pm 1,1}= \sqrt{3/4\pi} \sqrt{\Sigma_{\ell}}\sqrt{\Sigma_{\ell\pm 1}}$
\ben
&& \tilde\Psi^{\parallel}_{\ell, m=0}(k\hat z) = 4\pi \sqrt {1 \over 3}\; 
{\tilde\bq}_{\parallel}({k\hat z};r)
\sum_{\ell'} (-i)^{\ell'} I_{\ell\ell'1} \; j_{\ell'}( kr) \sqrt{\Sigma_{\ell'}}\left(  \begin{array}{ c c c }
     \ell' & \ell & +1 \\
     0 & 0 & 0
  \end{array} \right)^2.
\een
As before, due to the triangular inequality, only terms that contribute are $\ell'=\ell\pm 1$.
\ben
&& \tilde\Psi^{\parallel}_{\ell, m=0}(k\hat z; r)=   4\pi \sqrt {1 \over 3}
\tilde{\bq}_{\parallel}({k\hat z}; r)
\Big [ (-i)^{\ell+1} {I}_{\ell,\ell+1,1}j_{\ell+1}(kr)\sqrt{\Sigma_{\ell+1}}\left(  \begin{array}{ c c c }
     \ell+1 & \ell & +1 \\
     0 & 0 & 0
  \end{array} \right)^2 \nn \\
&& \quad\quad\quad\quad\quad\quad\quad\quad + (-i)^{\ell-1} {I}_{\ell,\ell-1,1}j_{\ell-1}( kr)\sqrt{\Sigma_{\ell-1}}\left(  \begin{array}{ c c c }
     \ell-1 & \ell & +1 \\
     0 & 0 & 0
  \end{array} \right)^2 \Big ].
\een
Using the expression in Eq.~(\ref{eq:new3j1}) and Eq.~(\ref{eq:new3j2}) for the $3j$ symbols that appear in the above expression we can write:
\ben
&& \tilde\Psi^{\parallel}_{\ell, m=0}(k\hat z; r) = - (-i)^{\ell+1}\sqrt{4\pi \over{\Sigma_{\ell}}} 
\tilde{\bq}_{\parallel}({k\hat z}; r) \left [ (\ell+1)j_{\ell+1}(kr) - \ell j_{\ell-1}(kr)\right ].
\een
Next we use the following recurrence relation for $j_{\ell}(x)$ to simplify further:
\ben
&& \ell j_{\ell-1}(x) - (\ell+1)j_{\ell+1}(x) = \Sigma_{\ell}\; {j^{\prime}_{\ell}(k\,r)}. 
\een
Here the prime denotes a derivative w.r.t. the argument. Using this expression we can write:
\ben
&& \tilde\Psi^{\parallel}_{\ell, m=0}(k\hat z)= {\sqrt{4\pi \Sigma_{\ell}}} \;
\tilde{\bq}_{\parallel}({k\hat z};r) \; {j^{\prime}_{\ell}(kr)}.
\een
For higher $\ell$ the contribution from $j_{\ell-1}$
and $j_{\ell+1}$ nearly cancels each other thus making ${\cal C}^{\parallel}_{\ell}(k)$ sub dominant.
\begin{figure}
\centering
\vspace{.25cm}
{\epsfxsize=16 cm \epsfysize=6 cm{\epsfbox[37 402 573 600]{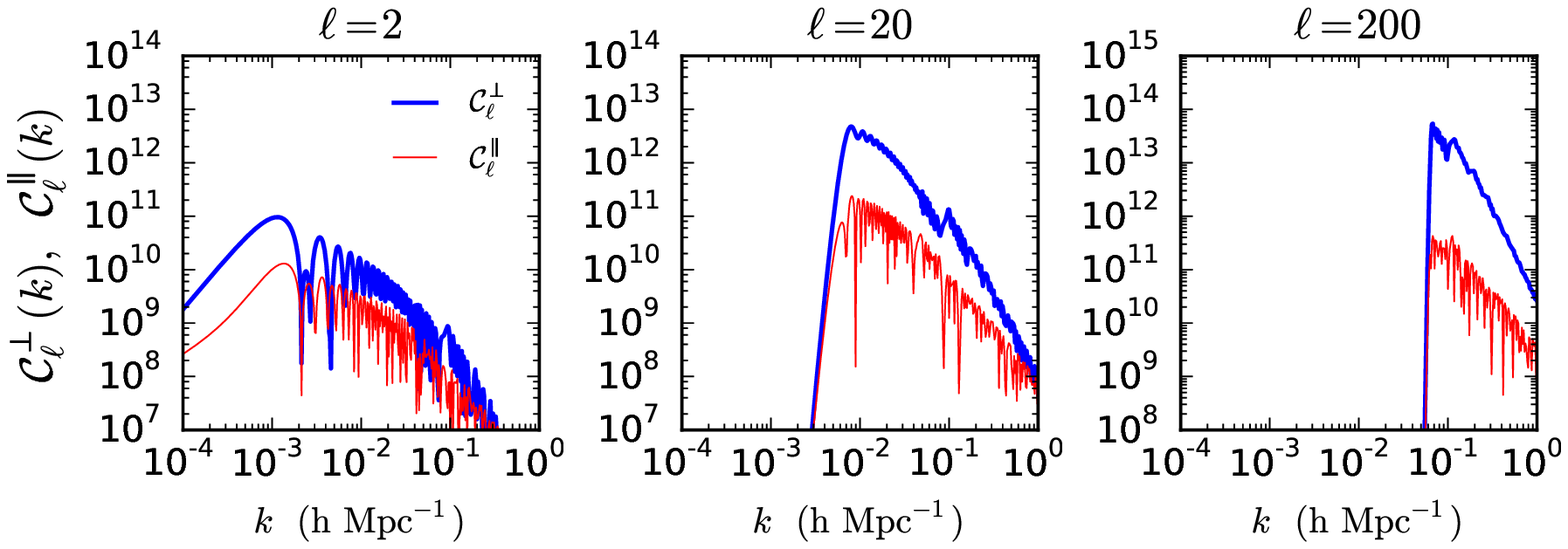}}}
\caption{The power spectra ${\cal C}^{\perp}_{\ell}(k)$ (defined in Eq.~(\ref{eq:factor_recon_perp1})) 
and ${\cal C}^{\parallel}_{\ell}(k)$ (defined in Eq.~(\ref{eq:2Dcls_para})) are displayed
for $\ell=2$ (left-panel), $\ell=20$ (middle-panel) and $\ell=200$ (right panel). The reconstruction was extended
upto a maximum rdshift of $z=1$. The power spectra are computed using underlying non-linear matter power spectra.}
\label{fig:cls2D}
\end{figure}
Performing the rotation using Wigner $D^{\ell}_{mm'}$ - matrices  as before to a general orientation we have:
\ben
&& \tilde\Psi_{\ell m}^{\parallel}(k) = (-i)^{\ell+1}\;\sqrt{4\pi\Sigma_{\ell}}\; \int{d^3 {\bf k^{\prime}} \over (2\pi)^3} 
\sum_m D^{\ell}_{m,0}({\bf \oh}_{k^{\prime}}) \int {dr\;r^2\; \phi(r)}\; k\; j_{\ell}(k\,r)\; j^{\prime}_{\ell}(k^{\prime}\,r)\; \tilde{\bq}_{\parallel}({k}'\hat z,r).
\een
Finally using the orthogonality relation for
Wigner D-matrices given in Eq.~(\ref{eq:d_matrix}), and the definition $P_{q\parallel}$ in Eq.~(\ref{eq:qdef}) we arrive at the following expression:
\ben
\;\;\;\;\;\;\;\; {\cal C}^{\parallel}_{\ell}(k_1,k_2)&\equiv& \la\hat\Theta_{\ell m}^{\parallel}(k_1)\hat\Theta_{\ell m}^{*\parallel}(k_2)\ra \nn \\
&=& 2\Pi_{\ell}\;\int \; dk\; k^2 \; \int d\,r_1\;r^2\; \phi(r_1)\; k_1 j_{\ell}(k_1r_1) {j^{\prime}_\ell(k\,r_1)}
\int d\,r_2\;r^2\; \phi(r_2)\; k_2\, j_{\ell}(k_2r_2)\; {j^{\prime}_\ell(k\,r_2)} \; P_{q\parallel}(k; r_1,r_2).
\label{eq:full_para}
\een
Consequently, we can use the approximation $P_{q\parallel}(k;\; r,r') \approx\sqrt{ P_{q\parallel}(k;\;r)}\sqrt{ P_{q\parallel}(k;\;r')}$
to factorize the integrals:
\ben
&&{\cal C}^{\parallel}_{\ell}(k_1,k_2) \approx 
\int\;  k^2 \; dk\; {\cal I}^{2\parallel}_{\ell}(k,k_1)\; {\cal I}^{2\parallel}_{\ell}(k,k_2);\label{eq:para_cl1}\\
&& {\cal I}^{2\parallel}_{\ell}(k,k')\approx \sqrt{2 \over \pi} \int \; dr \;r^2\; \phi(r) \;
k\, j_{\ell}(kr)\;r^2\; j^{\prime}_{\ell}(k^{\prime}\,r)\; \sqrt{P_{q\parallel}(k; r)}.
\label{eq:para_cl2}
\een
%
Cross-correlating the 2D CMB maps against reconstructed 3D estimators:
\ben
&&{\hat{\cal C}}^{\parallel}_{\ell}(k) \equiv \la\Theta_{\ell m}\hat\Theta_{\ell m}^{*\parallel}(k)\ra = 
2\Pi_{\ell}\;\int \; dk^{\prime}\; {k^{\prime}}^2 \; \int d\,r_1\;r_1^2 \phi(r_1)\; {j^{\prime}_\ell(k^{\prime}\,r_1)}
\int d\,r_2\;r_2^2\; \phi(r_2)\; k\, j_{\ell}(k\,r_2)\; {j^{\prime}_\ell(k^{\prime}\,r_2)} \; P_{q\parallel}(k^{\prime}; r_1,r_2).
\label{eq:full_recon_perp}
\een
The corresponding expressions using the factorization are as follows:
\ben
&&{\hat{\cal C}}^{\parallel}_{\ell}(k) \approx \int \; dk^{\prime}\; {k^{\prime}}^2 \; {\cal I}^{2\parallel}_{\ell}(k,k^{\prime})\; {\cal I}^{\parallel}_{\ell}(k^{\prime});\\
\label{eq:2Dcls_para}
&& {\cal I}^{\parallel}_{\ell}(k) \approx \sqrt{2 \over \pi}\int\; dr\; r^2 \;\phi(r)\; {j^{\prime}_{\ell}(kr)} \sqrt{P_{q\parallel}(k;r)}. 
\label{eq:2Dkernel_para}
\een
\label{sec:results}

Various approximations are typically used to collapse a full 3D analysis to 
projection (2D) or, in other words to perform tomography (see \textsection\ref{sec:tomo}). However, these methods invariably involve 
binning in redshift interval with a consequent loss of the information content of the data.
The 3D analysis also gives the freedom to treat the angular and radial
modes separately. Projection is achieved using Limber approximation invariably links the radial and angular modes.
In projected or tomographic survey its very difficult to disentangle individual scales.
Treatment of individual scales independently is important for keeping a handle on the effect of non-linearity
due to gravitational evolution.
A 3D analysis treats each galaxy individually and thus can be use to control the
systematics at a level of individual galaxies, which is very important for kSZ reconstruction.
Any contaminating region can always be masked out. A pseudo-${\cal C}_{\ell}$ (PCL) based
approach will be developed that can be used to recover the power spectra in the
presence mask (see Appendix-\ref{sec:mask}). 

\begin{figure}
\centering
\vspace{0.25cm}
{\epsfxsize=16 cm \epsfysize=6 cm{\epsfbox[37 402 573 591]{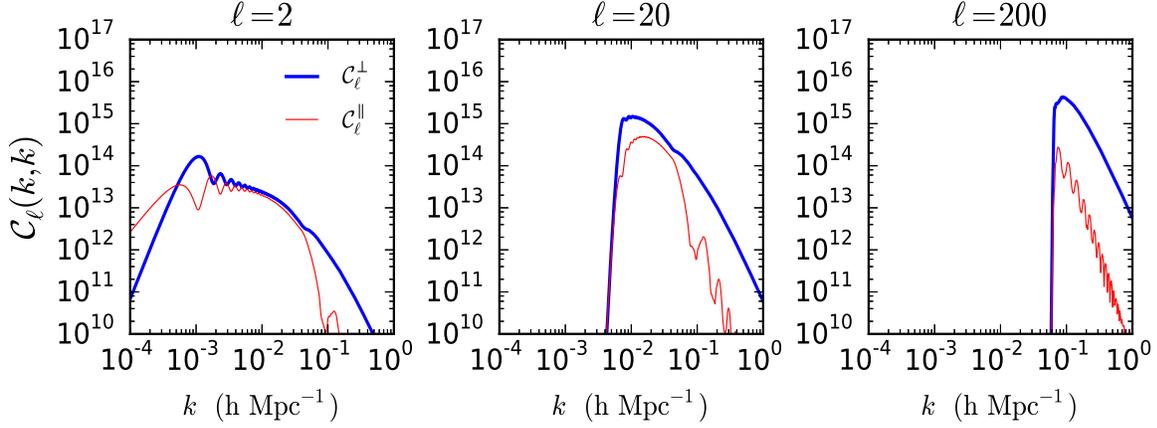}}}
\caption{The diagonal components of the 3D covariance matrix ${\cal C}^{\perp}_{\ell}(k,k')$ (defined in Eq.~(\ref{eq:perp_cl1})) 
and ${\cal C}^{\parallel}_{\ell}(k,k')$ (defined in Eq.~(\ref{eq:full_para})) are displayed
for $\ell=2$ (left-panel), $\ell=20$ (middle-panel) and $\ell=200$ (right panel). The reconstruction was extended
up to a maximum rdshift of $z=1$. The power spectra are computed using underlying non-linear matter power spectra.}
\label{fig:cls3D}
\end{figure}
\section{Projecting 3D onto Tomography and 2D}
\label{sec:tomo}
In order to connect with the tomographic reconstruction of kSZ effect by \citep{Shao11a},
in this section we will show how to reconstruct the tomographic power spectra from the 
3D power spectra we have developed.
The tomographic estimator $\hat\Theta^{\perp(i)}$ from the $i$-th bin as introduced in Eq.~(\ref{eq:theta_divide}) and its harmonic coefficients $\Theta^{\perp(i)}_{\ell m}$
can be expressed in terms of their 3D counterparts : 
\ben
&& \hat\Theta^{\perp(i)} = \int^{r_{i+1}}_{r_i} \, dr \, w(r)\, \hat\Theta^{\perp}({\bf r}); \\
&& \hat\Theta^{\perp(i)}_{\ell m} = \int\, d{\bf \oh}\,Y_{\ell m}({\bf \oh}) \sqrt{2 \over \pi} \sum_{\ell^{\prime}m^{\prime}} 
\int dk\, k\, j_{\ell}(kr)\, Y_{\ell^{\prime} m^{\prime}}({\bf\oh})\, \Theta^{\perp}_{\ell^{\prime} m^{\prime}}(k).
\een
Using the orthogonality property of spherical harmonics we can express the  $\Theta^{\perp(i)}_{\ell m}$
in terms of $\Theta^{\perp}_{\ell m}(k)$ using a kernel $T^{(i)}_{\ell}(k)$. The $i$-th bin is defined
between the radial distance $r_i$ and $r_{i+1}$.
\ben
&& \hat\Theta^{\perp(i)}_{\ell m} = \int \,dk\, {\cal J}^{(i)}_{\ell}(k)\hat\Theta^{\perp}_{\ell m}(k);  
\quad\quad T^{(i)}_{\ell}(k)= \sqrt{2 \over \pi}k\int^{r_{i+1}}_{r_i} dr\, j_{\ell}(k\,r)\, w(r).
\een
The tomographic power-spectra for the $i$-th bin is denoted as ${\cal C}^{\perp(i)}_\ell$.
Similarly the covariance between $i$-th and $j$-th bin is denoted as ${\cal C}^{\perp(ij)}_{\ell}$.
\ben
&&\hat{\cal C}^{\perp(ij)}_{\ell} \equiv \la \hat\Theta^{\perp(i)}_{\ell m}\hat\Theta^{\perp(j)*}_{\ell m} \ra = 
\int \,dk\, {\cal J}^{(i)}_{\ell}(k)\, \int \,dk' {\cal J}^{(j)}_{\ell}(k')\, {\cal C}_\ell^{\perp}(k,k'); \label{eq:cov}\\
&&\hat{\cal C}^{\perp(i)}_\ell \equiv \la\hat\Theta^{\perp(i)}_{\ell m}\Theta^*_{\ell m}\ra= \int\, dk\, {\cal J}^{(i)}_{\ell}(k)
{\cal C}_\ell^{\perp}(k).
\een
Similar expressions were derived by \cite{Kit14} in the context of 3D weak lensing.
An exactly similar relation can be obtained for ${\cal C}^{\parallel(ij)}_{\ell}$ and ${\cal C}^{\parallel(i)}_{\ell}$. To construct the 2D estimator
given in Eq.~(\ref{eq:2DCls}) we need to use only one bin and use the Limber approximation with $w=1$.

The tomography essentially probes discrete sets of physical wave numbers. The position of the tomographic
bins constraints the radial wave numbers probed as a function of angular harmonics.
The advantage of 3D is being able to ensure the integrated analysis of the entire set of wave-numbers
probed by a survey. Indeed, tomography allows for straightforward interpretation of the results
particularly redshift dependence of the kSZ signal. However, using Limber approximation as outlined
above can always be used to collapse the 3D information to tomographic bins.   
\section{Signal-to-Noise of Reconstruction}
\label{sec:s2n}
A simplistic expression for the signal-to-noise $(\rm S/N)$ for reconstruction of the individual modes described by radial and
angular wave numbers $(k,\ell)$ respectively is given by: 
\ben
\left ({S \over N}\right )^{\perp}_{\ell}(k) = f_{\rm sky}
\sqrt{2\ell+1 \over 2}{{\cal C}^{\perp}_{\ell}(k)\over\sqrt{ {\cal C}^{\perp}_{\ell}(k,k^{\prime}){\cal C}^{\rm CMB}_{\ell}
+ {\cal C}^{\perp}_{\ell}(k){\cal C}^{\perp}_{\ell}(k)}}.
\label{eq:s2n}
\een
Expressions for ${\cal C}^{\perp}_{\ell}(k,k^{\prime})$ is derived in Eq.~(\ref{eq:perp_cl1}) and ${\cal C}^{\perp}_{\ell}(k)$ in Eq.~(\ref{eq:factor_recon_perp2}).
The CMB power spectrum ${\cal C}^{\rm CMB}_{\ell}$ should also include the tSZ power spectrum at high-$\ell$ for $(\rm S/ N)$ estimation. 
We assume the noise of estimation to be independent of CMB noise; $f_{\rm sky}$ denotes the fraction of sky covered.
A similar expression for ${\cal C}^{\parallel}_{\ell}(k)$ can be obtained by replacing  ${\cal C}^{\perp}_{\ell}(k)$ and ${\cal C}^{\perp}_{\ell}(k)$
with ${\cal C}^{\parallel}_{\ell}(k)$ and ${\cal C}^{\parallel}_{\ell}(k)$.
\begin{figure}
\centering
\vspace{0.25cm}
{\epsfxsize=16 cm \epsfysize=6 cm{\epsfbox[37 402 573 591]{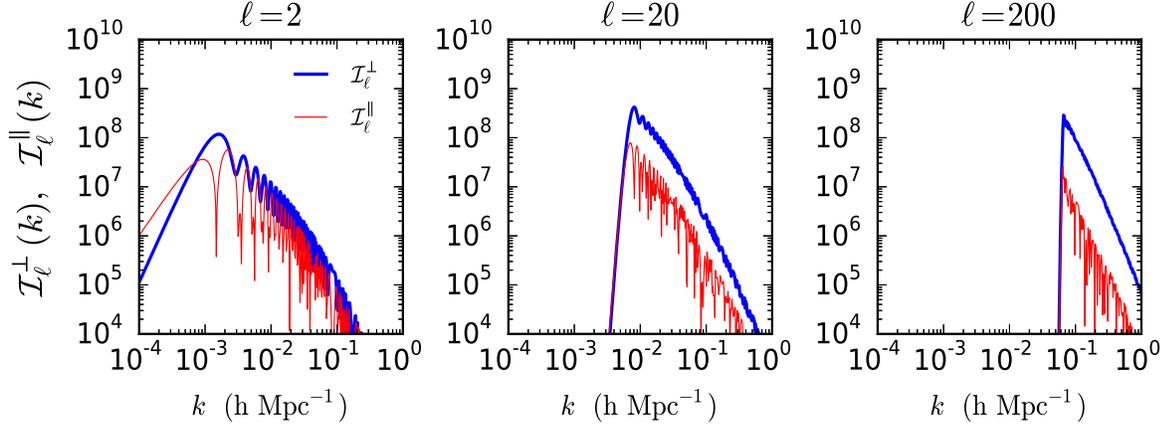}}}
\caption{The diagonal entries of the 2D kernels ${\mathcal I}^{\perp}_{\ell}(k)$ and ${\mathcal I}^{\parallel}_{\ell}(k)$
defined in Eq.~(\ref{eq:factor_recon_perp2}) and in Eq.~(\ref{eq:2Dkernel_para}) are plotted as a function of the wave number $k$ in $\rm h Mpc^{-1}$. The panels from left to right
correspond to $\ell=2, 20$ and $200$ respectively. Reconstruction is performed up to a maximum redshift of $z=1$.
Underlying matter power spectrum is assumed to be non-linear.}
\label{fig:kernel2D}
\end{figure}
\section{Input Power Spectra for $\tilde{\bq}_{\parallel}$ and $\tilde{\bq}_{\perp}$}
\label{sec:ps}
The computation of the kSZ power spectra depends on modelling $P_{q\perp}$.  Early analytical results were based on the linear perturbation theory and ignored non-linearity in density and velocity field \citep{Vis87}. 
The kSZ effect in linear regime is also known as the Ostriker-Vishniac (OV) effect \citep{OV86}. In \cite{Hu00} the linear 
theory prediction for the density power spectrum was replaced with non-linear prescription for density power spectrum by \cite{PD96}. 
The velocity power spectrum was from linear theory prediction (see e.g. \cite{DJ93} and  \cite{Hu00} for a detailed discussion of this issue). 
In \cite{MaFry02} the non-linear power spectrum used was that from halo-model prescription.
We outline the major step in the derivation to underline the approximations used in various steps.
In terms of density $\tilde\delta({\bf k})$ and velocity ${\bf v(k)}$ components we have: 
\ben
&& \tilde{\bq}_{\parallel}({\bf k}) = \int {d^3 {\bf k'} \over (2\pi)^3} \; \mu\, \hat{\bf \Omega}_k\; \tilde {\bf v}(k')\tilde \delta(|{\bf k-k'}|);\quad\\
&& \tilde{\bq}_{\perp}({\bf k}) = \int {d^3 {\bf k'} \over (2\pi)^3} (\hat{\bf\Omega}_{k'} - \mu \hat{\bf \Omega}_k) \tilde {\bf v}(k')\tilde\delta(|{\bf k-k'}|); \quad
\mu={\bf \oh}\cdot{\bf \oh}_{\rm k}.
\een
The corresponding 3D power spectra are given by:
\ben
&& P_{q\perp}(k,r) \equiv \la\tilde{\bq}_{\perp}({\bf k})\tilde{\bq}^*_{\perp}({\bf k}) \ra 
= \int {d^3 {\bf k'} \over (2\pi)^3}\left [ (1-{\mu^{\prime}}^2)  P_{\delta\delta}(|{\bf k-k^{\prime}}|)P_{vv}(k^{\prime}) - 
{(1-{\mu^{\prime}}^2)k^{\prime}\over |{\bf k-k^{\prime}}|} P_{\delta v}(|{\bf k-k^{\prime}}|)P_{\delta v}(k^{\prime})  \right ]; \label{eq:perp}\\
&& P_{q\parallel}(k,r)\equiv \la\tilde{\bq}_{\parallel}({\bf k})\tilde{\bq}^*_{\parallel}({\bf k}) \ra 
 =\int {d^3 {\bf k^{\prime}} \over (2\pi)^3} \left [ {\mu^{\prime}}^2 P_{\delta\delta}(|{\bf k-k^{\prime}}|)P_{vv}(k^{\prime}) 
+ {(k-k^{\prime}\mu^{\prime})\mu^{\prime} \over  |{\bf k-k^{\prime}}|} P_{\delta v}(|{\bf k-k^{\prime}}|)P_{\delta v}(k^{\prime})  \right ]. \label{eq:para}
\een
These expressions were computed by neglecting the connected fourth order moments which were found to be negligible.
We will use the following expression from \citet{Vis87} (see also \citet{JK98, MaFry02,Park15}) in our derivation of the 3D kSZ power spectrum: 
\ben
&& P_{q\perp}(k,r) = \dot a^2 f^2 \int {d^3{\bf k'} \over (2\pi)} P_{\delta\delta}(|k-k'|)P_{\delta\delta}(k') {k(k-2k'\mu)(1-\mu'^2) \over k'^2(k^2+k'^2 -2 k k'\mu')}. \label{kperp} \\
&& P_{q\parallel}(k,r) = \dot a^2 f^2 \int {d^3{\bf k'} \over (2\pi)} P_{\delta\delta}(|k-k'|)P_{\delta\delta}(k') 
{k\mu'(k \mu' -2 k'\mu'^2 +k' ) \over k'^2(k^2+k'^2 -2 k k'\mu') }; \quad\quad f \equiv {d \ln D_{+} \over d \ln a}. \label{kpara}
\een

The above expressions were derived under the assumption $P_{\delta \theta}(| k-k^{\prime}|)P_{\delta\theta}(k^{\prime})= P_{\delta\delta}(| k-k^{\prime}|) P_{\theta\theta}(k^{\prime})$
\citep[see e.g.][and references therein]{MF00}.
This approximation is a necessary ingredient as the density-velocity cross-correlation is poorly understood.
In the linear theory the the velocity and cross-correlation power spectra are expressed as follows:
\ben
&& P_{vv}(k) = \left({f \dot a\over k} \right )^2 P_{\delta\delta}(k); \quad\quad P_{\delta v}(k) = \left({f \dot a\over k} \right ) P_{\delta\delta}(k).
\label{eq:power_vv}
\een
To evaluate the kSZ power spectra defined in Eq.~(\ref{eq:factor_recon_perp1}) and Eq.~(\ref{eq:2Dcls_para}) we use 
these expressions for $P_{q\perp}(k,r)$ and $P_{q\parallel}(k,r)$.
Following \cite{MF00} we use the halo model to compute the density power spectra and linear theory to compute the
velocity power spectrum (see \cite{DJ93} and \cite{Hu00} for related discussions). In the high -k limit 
Eq.~(\ref{eq:perp})-Eq.~(\ref{eq:para}) is approximated as:
\ben
&& P_{q_{\perp}}(k) = {2 \over 3} P_{\delta\delta}(k)\;v^2_{\rm rms}(k); \quad\quad P_{q_{\perp}}(k)= 2P_{q_{\parallel}}(k); \quad\quad v^2_{\rm rms}(k) \equiv \int_{k^{\prime} \le k} {d^3k^{\prime} \over (2\pi)^3} P_{vv}(k^{\prime}).
\een
Here $v^2_{rms}$ is the velocity dispersion. In a recent study the effect of the four-point correlation 
function, which we have ignored in our study has been outlined by \citet{Park15}. They pointed out
that in the non-linear regime at $k>0.1 h\rm Mpc^{-1}$ the contribution from the four-point correlation
function (connected-part) can reach as high as $10\%$. 

\section{Results and Discussion}
In Figure~\ref{fig:pofk} we show the input power-spectra $P_{q}(k,z)$ (right-panel) $P_{q\perp}(k,z)$ (left-panel) and $P_{q\parallel}(k,z)$ (middle-panel)
used for our calculation. The solid lines represent
non-linear power-spectra and the dashed lines linear power-spectra. The results are for $z=0$. In general, to
relate the baryonic power spectrum and the dark matter power spectrum a window function is used. 
We have not included any such filtering in our computation to keep the analysis transparent, however such modification is straightforward.
Notice that the treatment we followed involves using the linear relationship in Eq.~(\ref{eq:lin_den_vel}) between density and velocity, which breaks down at high $k$. Accurate modelling in that regime can only be obtained using numerical simulations.
Nevertheless, we believe the qualitative nature of our results will remain unchanged in the range of $(k,\ell)$ considered here. 
The approximate form of the power spectrum we have chosen that allows us to separate two internal integrals and reduces 
computation time significantly. Indeed it can be argued that presence of spherical Bessel functions introduces
a cut-off that suppresses long range (low-$\ell$) modes thus justifying such a factorization.

In Figure~\ref{fig:cls2D} we show the computed ${\cal C}^{\perp}_{\ell}(k)$ and ${\cal C}^{\parallel}_{\ell}(k)$ as a function of $k$ for $\ell=2$, $\ell=20$ and $\ell=200$.  These spectra were defined in Eq.~(\ref{eq:perp_cl1}) and Eq.~(\ref{eq:factor_recon_perp1})
respectively. The spectra correspond to a maximum redshift of reconstruction of $z=1$. Different panels correspond
to harmonics $\ell=2,20$ and $200$ from left to right. The peak of the spectra shifts to higher $k$ with increasing $\ell$.  
Due to the presence of the spherical Bessel functions, which become increasingly sharp-peaked, the 3D cross-spectra
effectively collapse to the projected ones at higher $\ell$s, thus virtually linking the radial wave-number $k$ and the angular wave-number $\ell$.
The range of $k$ that contributes to a given $\ell$ becomes more sharply defined. 

The power spectrum  ${\cal C}^{\perp}_{\ell}(k)$ increasingly dominates over ${\cal C}^{\parallel}_{\ell}(k)$ 
with increasing $\ell$. In Figure~\ref{fig:cls3D} the diagonal entries for 3D covariance ${\cal C}^{\perp}_{\ell}(k,k)$ 
(solid-curves) and ${\cal C}^{\parallel}_{\ell}(k,k)$ (dashed-curves)
are shown as a function of wave number $k(h^{-1}\rm Mpc)$. These covariances are defined in Eq.~(\ref{eq:perp_cl1}) and Eq.~(\ref{eq:para_cl1}) respectively. These plots show diagonal cuts in the $(k,k^{\prime})$ plane for 
various $\ell$ values through the 3D covariance matrices. The effect of the Bessel function is to introduce 
a sharp cut-off on the long wave length contributions with $k < \ell / r_{max}$, where $r_{max}$ is the extent 
of the survey. As the $\ell$ values increases the diagonal entries terms of the covariance do not become 
significant until $k r_{max} \approx \ell$.  

\begin{figure}
\centering
\vspace{0.25cm}
{\epsfxsize=16 cm \epsfysize=6 cm{\epsfbox[37 402 573 591]{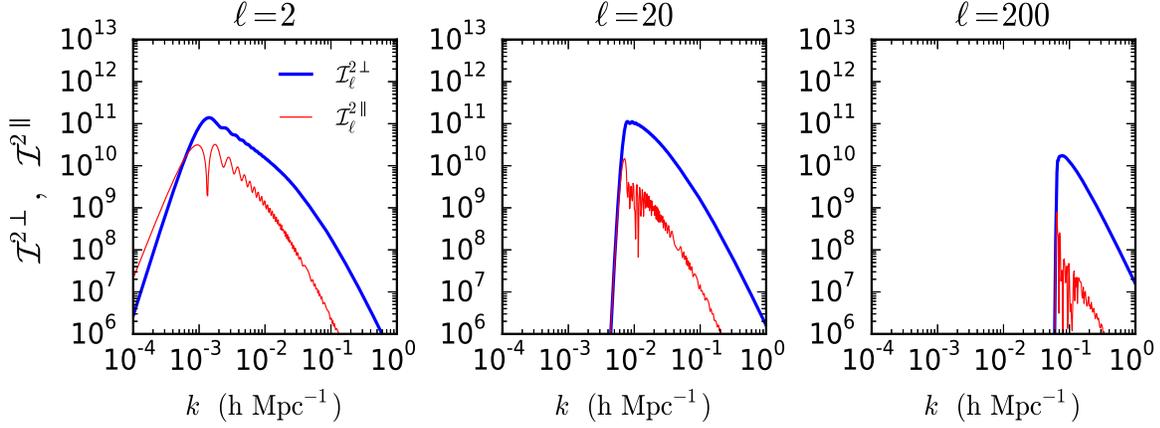}}}
\caption{The diagonal entries of the kernels ${\cal I}^{2\perp}_{\ell}(k,k')$ and ${\cal I}^{2\parallel}_{\ell}(k,k')$ 
defined in Eq.~(\ref{eq:perp_cl2}) and Eq.~(\ref{eq:para_cl2}) 
are plotted as a function of the wave number $k$ in $h^{-1}\rm Mpc$.}
\label{fig:kernel3D}
\end{figure}

The kernels ${\cal I}^{\perp}_{\ell}(k)$ and ${\cal I}^{\parallel}_{\ell}(k)$ which are 
used to construct the cross-spectra ${\cal C}^{\parallel}_{\ell}(k)$ (dashed-lines) and ${\cal C}^{\perp}_{\ell}(k)$ (solid-lines) 
 as a function of wavelength are plotted in Figure~\ref{fig:kernel2D}.
The corresponding kernels ${\cal I}^{2\perp}_{\ell}(k,k')$ and ${\cal I}^{2\parallel}_{\ell}(k,k')$ are shown in Figure~\ref{fig:kernel3D}. 

\begin{figure}
\centering
\vspace{0.25cm}
{\epsfxsize=5 cm \epsfysize=5 cm{\epsfbox[97 211 374 471]{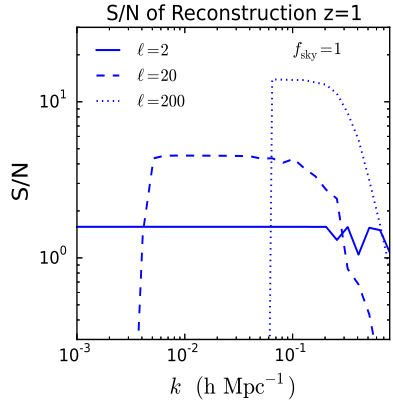}}}
\caption{We show an order of magnitude estimate of the signal to noise $\rm S/N$ introduced in Eq.~(\ref{eq:s2n}) for three modes with angular harmonics
represented by $\ell=2,20$ and $200$ as a function of their radial wave number $k (h^{-1}\rm Mpc)$ (see text for more details). We assume an
all-sky coverage $f_{\rm sky}=1$.}
\label{fig:s2n}
\end{figure}

In Figure~\ref{fig:s2n} we display the result of $({\rm S/N})$ computed using Eq.~(\ref{eq:s2n}). We show results for  $\ell=2,20$ and $200$
as a function of the radial wave number $k$. The reconstruction is assumed to be done up to a redshift of $z=1$. We ignore the residuals from
component separation. We also ignore the shot-noise due to finite number density of galaxies. However, the S/N of maps from galaxy surveys
are much higher compared to their CMB counterpart and are not likely affect the S/N of the reconstruction. Higher multipoles
have better $(\rm S/N)$ of reconstruction primarily due to the $2\ell+1$ prefactor. However, the available range of radial wavenumber $k$
decreases with increasing $\ell$. At higher $k$ the linear reconstruction of velocity from density field may not be accurate.  
Increasing the survey depth will clearly shift the peak of the reconstructed ${\cal C}_{\ell}(k)$ to lower radial wave-numbers $k$, but, number density of galxies
used for reconstruction will decrease thus decreasing the S/N. Maps from future 21cm surveys or other tracers of large scale structure can be used 
to extend the reconstruction technique proposed here to moderate redshifts $z<3$.
\section{Conclusions and Outlook}
\label{sec:conclu}
The kSZ signal is overwhelmed by various contaminations and lacks any redshift information due to 
the projection effect. Moreover, the contribution to kSZ can be from the reionization epoch as well as
from post-reionization era. The contribution from reionization depends on accurate
modelling of reionization which can in principle be inhomogeneous (patchy) and thus difficult
to model. With the recent detection of kSZ effect there is a greater effort in the community
to understand it than was previously. In this paper we focus on estimation of post-reionization
late-time kSZ effect. This contribution is relatively easier to model as the physics involve
is simpler compared to reionization and may provide us clues about our local universe and when
subtracted from the total kSZ can help constrain the contribution originating from reionization.

Indeed from CMB measurements alone, we only expect to detect kSZ from the auto power-spectra
of temperature fluctuations. However, kSZ is likely to dominate the power spectrum at $\ell>3000$
where the primary CMB is damped due to silk-damping and contribution from tSZ shows a null
at $\nu=217$ GHz. The method proposed here can probe kSZ below $\ell<3000$ as the 
method proposed here is independent of contamination from primary CMB or other secondary anisotropies.

The method proposed here uses spectroscopic galaxy surveys to reconstruct peculiar velocity field. 
Our method weighs the peculiar velocity field with galaxy distribution.
The resulting momentum when projected along the line of sight direction can act as an estimator
for kSZ effect in 3D. Our analysis extends previous analysis performed using tomographic bins \citep{HDS09,Shao11a}
and allows maximal exploitation of the available data with greater degree of freedom. 
We have constructed two different 3D estimators for kSZ power spectrum ${\mathcal C}^{\perp}_{\ell}(k)$ in Eq.~(\ref{eq:para_cl2})
and ${\mathcal C}^{\parallel}_{\ell}(k)$ in Eq.~(\ref{eq:perp_cl2}) that can be used to extract information.
This will eliminate any contamination from primary CMB which remains one of the
dominant source especially at low $\ell$. The tSZ contribution to CMB sky is typically
removed by using frequency information. The lensing of CMB and 
kSZ lacks any frequency information. In the past, several methods have been
proposed to separate such contribution, including the use of higher-order information or the non-Gaussianity of the signal. Most of the CMB secondary anisotropies originate in the
relatively low redshift universe and have a characteristic redshift evolution.
Using this fact we propose a cross-correlation based estimator to extract
kSZ effect from CMB maps. The proposal is based on using the spectroscopic
redshift surveys to construct an estimator for the peculiar velocity
and density. With suitable weighting these reconstructed fields can next be used to
construct an estimator for kSZ that takes into account the spin-1 nature of kSZ.
When cross-correlated with the CMB sky it will only cross-correlate with
the kSZ component thus avoiding usual contamination from other secondaries. 
There are many advantages to the technique developed here. It separates the 
late time, $(0<z<3)$ contribution to kSZ effect from the poorly understood {\em patchy reionization} scenarios.

The kSZ is effect is known to be a potentially powerful probe of the baryons
- the estimators presented here can exploit this to probe directly the 3D distribution of the missing baryons 
in the low redshift universe. The galaxy surveys typically have relatively higher $\rm (S/N)$ and are thus expected  
to allow such reconstruction. For tomographic reconstruction using Planck as target CMB survey
and BOSS as a spectroscopic galaxy survey a  $\rm (S/N)$ of $50$ was shown to be achievable.
It is expected that SZ survey e.g. SPT and galaxy surveys such as ADEPT, Euclid and SKA
would work still better. This will thus provide measurement of the power-spectra over a range of scales ($\ell$ and $k$) with a reasonable accuracy. The detection of baryons using the methods presented here using
kSZ signature in CMB maps does not rely on detection of 
hot gas or metals; it directly detects the free electrons in the IGM. The line-of-sight 
velocity is uncorrelated with many standard systematics that plague usual power spectrum analysis
applied to CMB data, e.g. it is insensitive to the unsubtracted tSZ, galactic foregrounds and detector noise. 
We have ignored the redshift space distortion in our formulation of an estimator for the kSZ.
The redshift space distortion can be effectively accounted for in the our 3D formulation
at least at scales where linear Kaiser effect remains valid. At smaller scales 
higher order contributions will have to be included. However at even smaller scales the finger-of-god effect caused by small scale random motion 
becomes important especially at low redshift where the perturbative results break down. Indeed our modelling of galaxy bias which we take as a linear and deterministic can also be 
stochastic \citep[see e.g.][and references therein]{Ahn15}. Such issues can only be dealt with through full numerical simulations.

The 3D reconstruction technique developed here is more fundamental than the tomographic
methods presented previously. Indeed the tomographic or the projected power spectra can both be constructed
from the 3D power spectra presented here. The 3D sFB transforms developed here
generalises similar expansion schemes used in other areas in cosmology including
e.g. studies of weak lensing \citep{Castro05} and redshift space distortions \citep{PM13}.

We have already emphasised the power of kSZ to map out the baryonic web. 
Constraints on the dark sector, i.e. decaying dark matter or dark matter-dark energy interactions
has recently attracted a lot of attention. They can be probed by tSZ or kSZ effects \citep[e.g.][]{Xu13}.
The power spectra $P_{q\perp}$ and $P_{q\parallel}$ depend on the underlying matter power spectra $P_{\delta}(k)$ as well as on the growth rate of perturbations, $f$, see Eq.~(\ref{kperp})-Eq.~(\ref{kpara}). Clearly
in many modified gravity theories such as the $f(R)$ theories, the Dilaton theories and
the K-mouflage models the power spectrum as well as the function $f(z)$ are very different 
from their $\Lambda$CDM counterparts. Future surveys with accurate determination
of the kSZ power spectrum may be able to constrain such theories.
Such issues in the context of 3D kSZ that we have studied here will be presented elsewhere.  
\section{Acknowledgements}
\label{acknow}
DM, ITI, KLD and PC acknowledge support from the Science and Technology
Facilities Council (grant number ST/L000652/1). The power spectrum $P_{q\perp}(k)$ and $P_{q\parallel}(k)$ used in
this work was computed by G. W. Pettinari. We acknowledge useful discussion with H. Park and P. Zhang. DM also acknowledges
results obtained by G. Pratten at the early stage of this work which helped obtaining some of the results presented here.
DM benefited from discussions with A. Heavens and P. Valageas.

\appendix
\section{Reconstruction in a Partial Sky}
\label{sec:mask}
The power-spectra we have constructed so far assume a full sky coverage. 
In general surveys will not cover the entire sky. However, once survey configuration
is specified by angular mask and selection function it is possible to use
a pseudo-${\cal C}_\ell$ (PCL) formalism in 3D to relate the observed ${\cal C}_{\ell}$s with the underlying
3D ${\cal C}_{\ell}$s. For the analytical derivation we will
assume that the 3D mask $ w_{\rm 3D}({\bf r})$ can be separated angular mask $w_a({\bf\oh})$ and the 
radial selection $w_r(r)$ can be treated separately. 
Denoting the masked field by $\Theta^{s\perp}({\bf r})$ which is recovered from a survey and the underlying
by $\Theta^{\perp}({\bf r})$ we can write:
\ben
&& \Theta^{s\perp}({\bf r}) = w_{\rm 3D}({\bf r})\Theta^{\perp}({\bf r}). 
\een
We will assume a separable mask $w_{\rm 3D}({\bf r})=w_r(r)w_a({\bf\oh})$.
This will allow us to deal with the angular and radial part separately making analytical treatment possible.
The harmonic coefficients of the masked field and the underlying field are related by the following expression:
\ben
&& \Theta^{s\perp}_{\ell m}(k) = \sum_{\ell^{\prime}m^{\prime}} \int \; dk^{\prime}\; \Theta^{\perp}_{\ell^{\prime} m^{\prime}}(k^{\prime}) W^{r}_{\ell\ell^{\prime}}(k,k^{\prime})
{\bar W}^a_{\ell m \ell^{\prime} m^{\prime}}.
\een
We will apply the same mask to the CMB $\Theta^{s\perp}(\oh)=w_a({\bf\oh})\Theta^{\perp}(\oh)$. The corresponding
angular harmonics denoted respectively as $\Theta^{s\perp}_{\ell m}$ and $\Theta^{\perp}_{\ell m}$ can be related as
\ben
&& \Theta^{s\perp}_{\ell m} =  \sum_{\ell^{\prime}m^{\prime}} {\bar W}^a_{\ell m \ell^{\prime} m^{\prime}}\Theta^{\perp}_{\ell^{\prime} m^{\prime}}.
\een
The radial and angular mode-mixing matrices denoted as  $W^r$ and $\bar W^a$ depends on the $w_r$ and $w_a$
respectively and can be expressed as follows:
\ben
&& W^r_{\ell\ell^{\prime}}(k,k^{\prime}) = {2 \over \pi}k\; k^{\prime}\;\int\; dr\; r^2 w_r^2(r) j_{\ell}(kr)j_{\ell^{\prime}}(k^{\prime}r);
\label{eq:window_radial} \\
&&  {\bar W}^a_{\ell\ell^{\prime}} = \int d{\bf \oh}\; w({\bf\oh})\; Y_{\ell m}({\bf \oh})\; Y_{\ell^{\prime} m^{\prime}}({\bf \oh}).
\label{eq:window_angular}
\een
The pseudo-${\cal C}_{\ell}$s for the masked field $\Theta^{s\perp}(\br)$ are denoted as ${\mathcal C}^{s\perp}_{\ell}(k)$
and ${\cal C}^{s\perp}_{\ell}(k_1,k_2)$. The pseudo ${\cal C}_{\ell}$s are linear combinations of the underlying ${\cal C}_{\ell}$s.
\ben
&& {\mathcal C}^{s\perp}_{\ell}(k) \equiv \la \Theta^{s\perp}_{\ell m}(k)\Theta^{s\perp *}_{\ell m} \ra= \sum_{\ell^{\prime}} W^a_{\ell\ell^{\prime}}\int dk^{\prime}
W^r_{\ell\ell^{\prime}}(k,k^{\prime})
{\mathcal C}^{\perp}_{\ell^{\prime}}(k^{\prime}); \label{eq:cross}\\
&& {\cal C}^{s\perp}_{\ell}(k_1,k_2) \equiv \la \Theta^{s\perp}_{\ell m}(k)\Theta^{s\perp *}_{\ell m}(k) \ra
= \sum_{\ell^{\prime}} W^a_{\ell\ell^{\prime}}\int dk^{\prime}_1\int dk^{\prime}_2 \; 
W^r_{\ell\ell^{\prime}}(k_1,k^{\prime}_1)\; W^r_{\ell\ell^{\prime}}(k_2,k^{\prime}_2)
{\cal C}^{\perp}_{\ell^{\prime}}(k^{\prime}_1,k^{\prime}_2) \label{eq:auto}.
\een
The linear transform relating the two power-spectra e.g. ${\mathcal C}^{s\perp}_{\ell}(k)$ and  ${\mathcal C}^{\perp}_{\ell}(k)$,
or, equivalently ${\cal C}^{s\perp}_{\ell}(k_1,k_2)$ and ${\cal C}^{\perp}_{\ell}(k_1,k_2)$, are encapsulated in a mode-mixing matrix.
However, due to our assumption of factorizability of the mask we can write the mixing matrix
as product of two different mode-mixing matrices.
The radial and angular mode-mixing matrices denoted as  $W^r$ and $W^a$ depends on the $w_r$ and $w_a$
respectively and can be expressed as follows:
\ben
&& {W}^a_{\ell\ell^{\prime}} = \sum_{L} w_{L} {I^2_{\ell L\ell^{\prime}} \over 2\ell+1} \left(  \begin{array}{ c c c }
     \ell & L & \ell^{\prime} \\
     0 & 0 & 0
  \end{array} \right)^2; \quad\quad w_{L} = {1 \over 2L+1}\sum_{M} w_{LM}w^*_{LM}.
\label{eq:coupling}
\een
For nearly all-sky coverage the mixing matrices can be inverted to recover the underlying ${\cal C}_{\ell}$s.
However, binning may be required for partial sky coverage.
The power-spectra ${\cal C}^{\parallel}_{\ell}(k)$ and ${\cal C}^{\parallel}_{\ell}(k,k^{\prime})$ can be dealt
with in an exactly similar manner. The mixing matrices remain unchanged.

It's possible to further simplify the expressions by using an approximate form for $j_{\ell}(x)$ valid at high $\ell$ \citep{LoAf08}:
\ben
&&  j_{\ell}(x)|_{\ell \rightarrow \infty}\approx \sqrt{\pi \over 2\ell +1 }\delta_D(\ell+{1 \over 2}-x).
\een

A few comments are in order. The PCL based sub-optimal estimators for projected or 2D data sets were developed by \cite{Hiv02} in the context of
CMB power spectrum estimation. It has also been generalised for computation of cross-spectra involving external
data sets \citep{MHCV}. More recently, it has been extended for the 3D data by \citep{MuKitch11} and was successfully
used for analysing 3D weak lensing power spectrum using CFHTLenS\footnote{\url http://www.cfhtlens.org/} data \citep{Kitch14}.
Though PCL based estimators remain sub-optimal, suitable use of weights can improve their performance.

The results presented here are generalisation of the one presented in \cite{MuKitch11} (where the effect of radial selection
function was not included; i.e. $W^r=1$.) and will have general applicability
in cross-correlating arbitrary projected field (3D) against a 3D cosmological tracer. Similar results can be derived
for higher-order spin fields by suitably modifying $W^{a}_{\ell\ell^{\prime}}$.

{\bf Generalisation to arbitrary spin:} For the sake of completeness we also provide the expression for a generic 3D field with an {\em arbitrary} spin $S$ $\alpha$.
We would like to cross-correlate it with another arbitrary projected field with spin $S^{\prime}$ say $\beta$. In this case
the basis of expansion involves spinorial harmonics ${}_SY_{\ell m}(\oh)$. The spin-harmonics are defined in
terms of $D$ matrices as $_{S}Y_{\ell m}({\bf\oh})\equiv \sqrt{2\ell+1 \over 4\pi} D^{\ell}_{-S,m}(\theta,\phi,0)$ \citep{PenRind,Var88}.
The forward and backward transformations
are obtained by replacing $Y_{\ell m}({\bf \oh})$ in Eq.~(\ref{eq:forward})-Eq.~(\ref{eq:backward}) with spin-weighted spherical harmonics
or spinorial harmonics ${}_SY_{\ell m}({\bf \oh})$:
\ben
&& {}_S\alpha^{s}_{\ell m}(k) = \sum_{\ell^{\prime}m^{\prime}} \int \; dk^{\prime}\; \alpha_{\ell^{\prime} m^{\prime}}(k^{\prime}) W^{r}_{\ell\ell^{\prime}}(k,k^{\prime})
{}_S{\bar W}^a_{\ell m \ell^{\prime} m^{\prime}}; \\
&& {}_S\beta^{s}_{\ell m} =  \sum_{\ell^{\prime}m^{\prime}} {}_S{\bar W}^a_{\ell m \ell^{\prime} m^{\prime}}\beta_{\ell^{\prime} m^{\prime}}.
\een
We follow the same notation as before but we assume the 3D field $\alpha({\bf r})$ is a spinorial field of spin-weight $S$ and 
$\beta({\bf \oh})$ is a projected filed of spin $S^{\prime}$. The harmonic components of these fields are now denoted as ${}_S\alpha_{\ell m}$
and ${}_{S^{\prime}}\beta_{\ell m}$. The suffix $s$ denotes quantities defined in partial sky.
The radial component of the mixing matrix given in Eq.~(\ref{eq:window_radial}) remains unchanged. However, the angular mode-mixing matrix depends on the spin
of the specific field Eq.~(\ref{eq:window_angular}).
\ben
&& {}_S{\bar W}^a_{\ell\ell^{\prime}} = \int d{\bf \oh}\; w({\bf\oh})\; {}_SY_{\ell m}({\bf \oh})\; {}_{S}Y^*_{\ell^{\prime} m^{\prime}}({\bf \oh}).
\een
The corresponding auto- and cross-spectrum are defined as follows:
\ben
&& {\mathcal C}^s_{\ell}(k) \equiv \la \alpha^{s}_{\ell m}(k)\beta^{s*}_{\ell m} \ra= \sum_{\ell^{\prime}} {}_{SS^{\prime}}W^a_{\ell\ell^{\prime}}\int dk^{\prime}
W^r_{\ell\ell^{\prime}}(k,k^{\prime})
{\mathcal C}_{\ell^{\prime}}(k^{\prime});\\
&& {\cal C}^{s}_{\ell}(k_1,k_2) \equiv \la \alpha_{\ell m}(k)\alpha^*_{\ell m}(k) \ra
= \sum_{\ell^{\prime}} {}_{SS^{\prime}}W^a_{\ell\ell^{\prime}}\int dk^{\prime}_1\int dk^{\prime}_2 \; 
W^r_{\ell\ell^{\prime}}(k_1,k^{\prime}_1)\; W^r_{\ell\ell^{\prime}}(k_2,k^{\prime}_2)
{\cal C}_{\ell^{\prime}}(k^{\prime}_1,k^{\prime}_2).
\een
These expressions generalises the ones obtained in Eq.~(\ref{eq:cross}-\ref{eq:auto}).
The coupling matrix is denoted by ${}_{SS^{\prime}}{W}^a_{\ell\ell^{\prime}}$ which is a generalisation
of the coupling matrix given in Eq.~(\ref{eq:coupling}):
\ben
&& {}_{SS^{\prime}}{W}^a_{\ell\ell^{\prime}} = \sum_{L} w_{L} {I^2_{\ell L\ell^{\prime}} \over 2\ell+1} \left(  \begin{array}{ c c c }
     \ell & L & \ell^{\prime} \\
     S & 0 & -S
  \end{array} \right)\left(  \begin{array}{ c c c }
     \ell & L & \ell^{\prime} \\
     S^{\prime} & 0 & -S^{\prime}
  \end{array} \right)  ; \quad\quad w_{L} = {1 \over 2L+1}\sum_{M} w_{LM}w^*_{LM}.
\een
Previous results correspond to the specific case when we set $S=S^{\prime}=0$.
One immediate practical use of the above expressions  could be joint analysis of CMB lensing and 3D galaxy lensing \citep{KHD15}.  
\end{document}